\documentclass[12pt]{article}

\usepackage[margin=1in]{geometry}

\usepackage{setspace}
\usepackage{floatrow}
\usepackage{lscape}
\usepackage{bbm}

\usepackage{amsmath, amssymb, amsfonts, amsthm}
\usepackage{bm}

\usepackage{graphicx}
\usepackage{booktabs}
\usepackage{threeparttable}
\usepackage{caption}
\usepackage{subcaption}

\usepackage[hidelinks]{hyperref}

\usepackage[
  backend=biber,
  style=apa,
  natbib=true,
  uniquename=false,
  doi=true,
  url=false,
  isbn=false,
  maxcitenames=2,
  maxbibnames=99
]{biblatex}
\DeclareLanguageMapping{english}{english-apa}

\usepackage{enumitem}
\setlist[itemize]{noitemsep, topsep=0pt}
\setlist[enumerate]{noitemsep, topsep=0pt}

\usepackage{microtype}

\usepackage{authblk}
\usepackage{footmisc}

\title{Sign-Dependent Spillovers of Global Monetary Policy}

\author{
Santiago Camara\thanks{Department of Economics, McGill University. Email: santiago.camara@mcgill.ca. 855 Sherbrooke Street West. Office L439. Leacock Building. Montreal (Quebec) H3A 2T7 Canada}
}

\date{\today}

\begin{document}

\maketitle

\begin{abstract}
This paper examines the sign-dependent international spillovers of Federal Reserve and European Central Bank monetary policy shocks. Using a consistent high-frequency identification of pure monetary policy shocks across 44 advanced and non-advanced economies and the methodology of \cite{caravello2024disentangling}, we document strong asymmetries in international transmission. Linear specifications mask these effects: contractionary shocks generate large and significant deteriorations in financial conditions, economic activity, and international trade abroad, while expansionary shocks yield little to no measurable improvement. Our results are robust across samples, identification strategies, and the framework proposed by \cite{ben2023government}.

\bigskip


\bigskip

\end{abstract}

% -------------------------------------------------
% Introduction
% -------------------------------------------------
\newpage
\section{Introduction}
\label{sec:introduction}

Monetary policy decisions by the world’s largest central banks have consequences that extend well beyond national borders. A large empirical literature documents that monetary policy shocks originating in the United States and, to a lesser extent, the euro area generate sizable international spillovers to exchange rates, asset prices, capital flows, and real economic activity. These spillovers are of first-order importance for small open economies whose financial conditions and business cycles are strongly influenced by global forces. Yet most existing empirical work relies on linear frameworks that impose symmetry between expansionary and contractionary monetary policy shocks.

This paper shows that this symmetry assumption is strongly rejected by the data. Using a unified high-frequency identification of pure monetary policy shocks and a sign-dependent local projection framework, we document that international monetary spillovers are overwhelmingly driven by contractionary shocks. For both the Federal Reserve and the European Central Bank, monetary policy tightenings lead to a pronounced deterioration in global financial conditions, significant contractions in real activity, and sharp declines in international trade. In contrast, monetary policy easings do not generate a symmetric reversal of these effects. Across many outcomes, responses to easing shocks are economically small, statistically insignificant, or even contractionary. As a result, linear specifications substantially understate the magnitude of international spillovers associated with monetary policy tightenings and, in some cases, misrepresent their direction.

Our empirical analysis follows the sign-dependent local projection framework proposed by \citet{caravello2024disentangling}, which provides a parsimonious benchmark specification to study asymmetries between tightening and easing shocks while preserving statistical power. In this framework, impulse responses are parameterized as a function of the monetary policy shock and its absolute value, allowing responses to differ by sign without imposing size nonlinearities. We adopt this specification as our benchmark and show that the resulting asymmetries are robust to alternative parameterizations of sign dependence, including the interacted local projection approach of \citet{ben2023government}.

We focus on the international transmission of monetary policy shocks originating in the United States and the euro area over the period from 1999 to 2023. Exogenous monetary policy shocks are identified using high-frequency movements in interest rate derivatives around policy announcements, purged of information effects following the decomposition approach of \citet{jarocinski2020deconstructing} and \cite{jarocinski2022central}. This identification strategy allows us to construct comparable monetary policy shock series for both central banks and to isolate the pure monetary policy component of policy announcements. Building on this identification, we estimate sign-dependent impulse responses using local projections.

The empirical strategy deliberately adopts a reduced-form perspective. Rather than imposing a structural model or privileging a particular transmission mechanism, we document how a broad set of observable macroeconomic, financial, and trade variables respond asymmetrically to monetary policy shocks of different signs. International spillovers operate through multiple, potentially interacting channels that are difficult to disentangle empirically without strong modeling assumptions. The contribution of this paper is therefore to provide a comprehensive and robust characterization of sign-dependent international spillovers, rather than to rank or quantify individual transmission channels.

A key implication of our findings is that linear empirical models provide a misleading characterization of international monetary transmission. By averaging strong contractionary spillovers with weak or countervailing responses to easing shocks, linear specifications attenuate estimated effects and, in some cases, obscure their sign. Allowing for sign dependence reveals that the international effects of monetary policy are not well approximated by a symmetric propagation mechanism. Importantly, our results are robust to alternative local projection designs that allow for state-dependent dynamics, including the approach of \citet{ben2023government}, confirming that the documented asymmetries do not hinge on a particular empirical specification.

\medskip

\paragraph{Related literature.} 
This paper relates to a large empirical literature studying the international spillovers of monetary policy shocks originating in the United States and the Euro Area. Recent work exploits high-frequency identification strategies to isolate exogenous monetary policy shocks and shows that U.S. monetary policy tightenings are associated with global financial tightening, currency depreciation, and contractions in output and trade abroad \citep{rey2016international, miranda2020us, camara2025spillovers, camara2025international}. Parallel work studies international spillovers from euro area monetary policy, documenting significant effects on financial conditions and real activity outside the euro area, particularly during periods of unconventional policy \citep{miranda2022tale, jarocinski2022central, camara2023international}. Our contribution to this literature is to show that, for both central banks, these international spillovers are highly asymmetric and overwhelmingly driven by contractionary shocks, a feature that is obscured in standard linear empirical frameworks.

The paper also contributes to a growing literature that emphasizes the nonlinear and asymmetric effects of macroeconomic policy shocks. A number of studies document that fiscal and monetary policy shocks can have state-dependent or sign-dependent effects, with expansionary and contractionary shocks generating markedly different responses \citep{ramey2018government,barnichon2022effects,ben2023government}. Recent contributions develop econometric frameworks that allow for asymmetries in impulse responses without imposing strong nonlinear structure, showing that policy tightenings often have larger and more persistent effects than easings \citep{caravello2024disentangling, ben2023government}. Our paper builds on this approach by applying sign-dependent local projections to the international transmission of monetary policy. We show that asymmetry is a central feature of global spillovers for both the Federal Reserve and the ECB, and that accounting for sign dependence is essential to accurately characterize the magnitude and direction of international monetary transmission.\footnote{\cite{camara2025between} studies sign-dependent monetary policy shocks of both the Federal Reserve and the ECB in the context of Canada, as part of a set of robustness exercises.}

\bigskip
\paragraph{Organization.}
The remainder of the paper is organized as follows. Section~\ref{sec:data_sample_methodology} describes the data, identification strategy, and econometric framework. Section~\ref{sec:sign_dependent_spillovers} presents the main results, contrasting linear and sign-dependent international spillovers for U.S. and euro area monetary policy shocks. Section~\ref{sec:understanding_asymmetry} documents how financial conditions and international trade respond asymmetrically to tightening and easing episodes. Section~\ref{sec:additional_results_robustness} provides an extensive set of robustness checks. The final section concludes.

%%%%%%%%%%%%%%%%%%%%%%%%%%%%%%%%%%%%%%%%%%%%%%%%%%%%%%%%%%%%%%%%%%%%%%
%%%%%%%%%%%%%%%%%%%%%%%%%%%%%%%%%%%%%%%%%%%%%%%%%%%%%%%%%%%%%%%%%%%%%%
\section{Sample, Identification Strategy \& Methodology} \label{sec:data_sample_methodology}

This section describes the data, identification strategy, and econometric methodology used in the empirical analysis. We first present the country sample and data construction. We then describe the high-frequency identification strategy used to isolate Federal Reserve and ECB monetary policy shocks purged of information effects. Finally, we outline the econometric framework used to estimate sign-dependent impulse responses and quantify the international
transmission of monetary policy shocks.

%%%%%%%%%%%%%%%%%%%%%%%%%%%%%%%%%%%%%%%%%%%%%%%%%%%%%%%%%%%%%%%%%%%%%%
%%%%%%%%%%%%%%%%% Sample
\subsection{Country Sample}
\label{subsec:sample}

Our analysis covers both advanced and non-advanced economies in order to provide a comprehensive assessment of the international transmission of Federal Reserve and ECB monetary policy shocks. We construct an unbalanced monthly panel spanning January 1999 to December 2023, starting with the inception of regular ECB monetary policy meetings.\footnote{In Section \ref{sec:additional_results_robustness}, we show that the results are robust to extending the sample for the Federal Reserve further back in time.}

We begin by assembling a benchmark dataset of advanced and non-advanced economies, defined according to the IMF classification. The benchmark sample includes all countries for which data are available on four core macroeconomic variables: (i) industrial production; (ii) consumer prices; (iii) the nominal bilateral exchange rate against the US dollar or the euro, depending on the monetary policy shock considered; and (iv) an equity price index. These series are obtained from the IMF’s International Financial Statistics and the OECD databases. Countries that later adopted the euro are retained in the sample until the date of euro adoption. This procedure yields a benchmark sample of 19 advanced economies and 25 non-advanced economies. The list of countries and the number of monthly observations available for each country are reported in Tables \ref{tab:sample_advanced} and \ref{tab:sample_non_advanced} in Appendix \ref{sec:appendix_data_details}.

To study transmission channels, we augment the benchmark dataset by adding one additional variable at a time. These variables include trade flows, commodity prices, domestic lending rates, policy interest rates, equity prices, sovereign bond spreads, international capital flows, unemployment, construction activity, government bond yields, and real effective exchange rates. This sequential augmentation strategy preserves the size of the country sample while allowing us to isolate specific channels through which Federal Reserve and ECB monetary policy shocks propagate internationally.

All specifications include US and Euro Area macro-financial control variables. For Federal Reserve shocks, we control for US industrial production, the PCE price index, the excess bond premium \cite{gilchrist2012credit}, and the federal funds rate. For ECB shocks, we control for the euro-area industrial production index, the harmonized consumer price index, a high-yield corporate bond spread, and the ECB policy rate. These controls account for contemporaneous
economic and financial conditions in the United States and the Euro Area that may independently affect macroeconomic outcomes abroad.

%%%%%%%%%%%%%%%%%%%%%%%%%%%%%%%%%%%%%%%%%%%%%%%%%%%%%%%%
\subsection{Identification Strategy}
\label{subsec:identification_strategy}

We identify exogenous monetary policy shocks for both the Federal Reserve and the European Central Bank using high-frequency identification techniques that exploit financial market reactions in narrow windows around policy announcements. Our benchmark identification follows the event-study and decomposition framework of \citet{jarocinski2020deconstructing}, which separates pure monetary policy shocks from information effects revealed at announcement times. This approach provides a unified and widely used identification strategy for both central banks, allowing for a consistent comparison of their international transmission.

For each central bank, identification is based on high-frequency surprises in interest rate derivatives measured within a 30-minute window around policy announcements and press conferences. These surprises capture unexpected changes in the perceived stance of monetary policy and are standard in the literature \citep{kuttner2001monetary, gertler2015monetary}. For the Federal Reserve, the shocks are constructed from U.S. interest rate futures, while for the ECB they are drawn from the Euro Area Monetary Policy Event-Study Database \citep{altavilla2019measuring}. In both cases, we source the monetary policy shock series directly from the authors’ publicly available replication files.

A key challenge in high-frequency identification is that monetary policy announcements may simultaneously convey information about the central bank’s assessment of economic conditions. Following \citet{jarocinski2022central}, we address this issue by exploiting the joint behavior of interest rates and equity prices around announcements. Pure monetary policy shocks are identified as innovations that raise interest rate expectations while lowering equity prices, whereas the orthogonal component captures information effects.\footnote{Operationally, we extract the first principal component of high-frequency interest rate surprises and apply a rotation-based sign-restriction procedure to isolate the monetary policy component. Because multiple admissible rotations satisfy the identifying restrictions, we follow the literature and focus on the median admissible rotation angle. Alternative identification strategies, including the Poor Man’s sign-restriction classification of \citet{jarocinski2020central}, yield similar results and are discussed in the robustness section.}

Throughout the paper, we take the identified Federal Reserve and ECB monetary policy shocks as given and focus on quantifying their international transmission. In Section~\ref{subsec:additional_identification}, we show that our results are robust to alternative high-frequency identification strategies, confirming that the documented asymmetries do not hinge on a particular shock construction.

%%%%%%%%%%%%%%%%% Econometric Specification
\subsection{Econometric Specification}
\label{subsec:econometric_specification}

We estimate impulse responses using local projections following \cite{jorda2005estimation}, augmented to allow for sign-dependent effects of monetary policy shocks as in \cite{caravello2024disentangling}. The empirical analysis in this paper focuses exclusively on sign dependence. Accordingly, the econometric specification includes the monetary policy shock and its absolute value, which is sufficient to capture asymmetries between contractionary and expansionary shocks under symmetry of the shock distribution.

For each country $i$ and horizon $h$, we estimate regressions of the form
\begin{equation}
y_{i,t+h}
=
\alpha_{i,h}
+
\beta_h^{\mathrm{sign}} \, \varepsilon_t
+
\beta_h^{\mathrm{abs}} \, |\varepsilon_t|
+
\Gamma_h X_{i,t-1}
+
u_{i,t+h},
\end{equation}
where $y_{i,t+h}$ denotes the outcome variable of interest, $\varepsilon_t$ is the externally identified monetary policy shock, and $X_{i,t-1}$ is a vector of lagged domestic and global controls. The specification includes country-by-calendar-month fixed effects, which absorb country-specific seasonal patterns and recurring calendar effects. This ensures that the estimated responses are not driven by systematic seasonality in macroeconomic outcomes, while preserving time variation needed to identify the effects of global monetary policy shocks. Standard errors are clustered at the time level to account for cross-country dependence induced by common global shocks.

The inclusion of the absolute value of the shock allows the response to depend on the sign of the monetary policy innovation. Under the assumption that the distribution of the shock is symmetric, the absolute value is an even function, and the coefficient $\beta_h^{\mathrm{abs}}$ isolates sign non-linearities rather than size effects \cite{caravello2024disentangling}. A rejection of the null hypothesis $H_0: \beta_h^{\mathrm{abs}} = 0$ therefore provides evidence of asymmetric responses to monetary policy tightenings and easings.

The estimated coefficients map directly into impulse responses for contractionary and expansionary shocks. The response to a contractionary (tightening) shock is given by
\begin{equation}
\mathrm{IRF}_h^{\mathrm{+}}
=
\beta_h^{\mathrm{abs}} + \beta_h^{\mathrm{sign}},
\end{equation}
while the response to an expansionary (easing) shock is given by
\begin{equation}
\mathrm{IRF}_h^{\mathrm{-}}
=
\beta_h^{\mathrm{abs}} - \beta_h^{\mathrm{sign}}.
\end{equation}
These expressions follow from the identities $|\varepsilon_t| = \varepsilon_t$ when $\varepsilon_t > 0$ and $|\varepsilon_t| = -\varepsilon_t$ when $\varepsilon_t < 0$.

Impulse responses for tightening and easing shocks are reported separately, with confidence intervals constructed using linear combinations of the estimated coefficients. Inference is conducted using standard local projection methods across horizons, with time-clustered standard errors.

By construction, this specification does not identify size non-linearities. The response is allowed to differ only by the sign of the monetary policy shock, reflecting the empirical focus of the paper on asymmetric international transmission of monetary policy tightenings and easings.

%%%%%%%%%%%%%%%%%%%%%%%%%%%%%%%%%%%%%%%%%%%%%%%%%%%%%%%%%%%%%%%%%%%%%%
%%%%%%%%%%%%%%%%%%%%%%%%%%%%%%%%%%%%%%%%%%%%%%%%%%%%%%%%%%%%%%%%%%%%%%
%%%%%%%%%%%%%%%%%%%%%%%%%%%%%%%%%%%%%%%%%%%%%%%%%%%%%%%%%%%%%%%%%%%%%%
%%%%%%%%%%%%%%%%%%%%%%%%%%%%%%%%%%%%%%%%%%%%%%%%%%%%%%%%%%%%%%%%%%%%%%
\section{Sign-Dependent Spillovers of Global Monetary Policy}
\label{sec:sign_dependent_spillovers}

This section presents the main empirical findings on the international spillovers of U.S. and euro area monetary policy shocks. We proceed in three steps. We first document the average (linear) spillover effects of monetary policy shocks. We then show that these average responses mask pronounced sign-dependent asymmetries for U.S. monetary policy shocks. Finally, we demonstrate that sign dependence is even more pronounced for euro area monetary policy, where international spillovers are overwhelmingly driven by contractionary shocks.

\medskip

\begin{figure}[ht]
    \centering
\includegraphics[width=14cm,height=10cm]{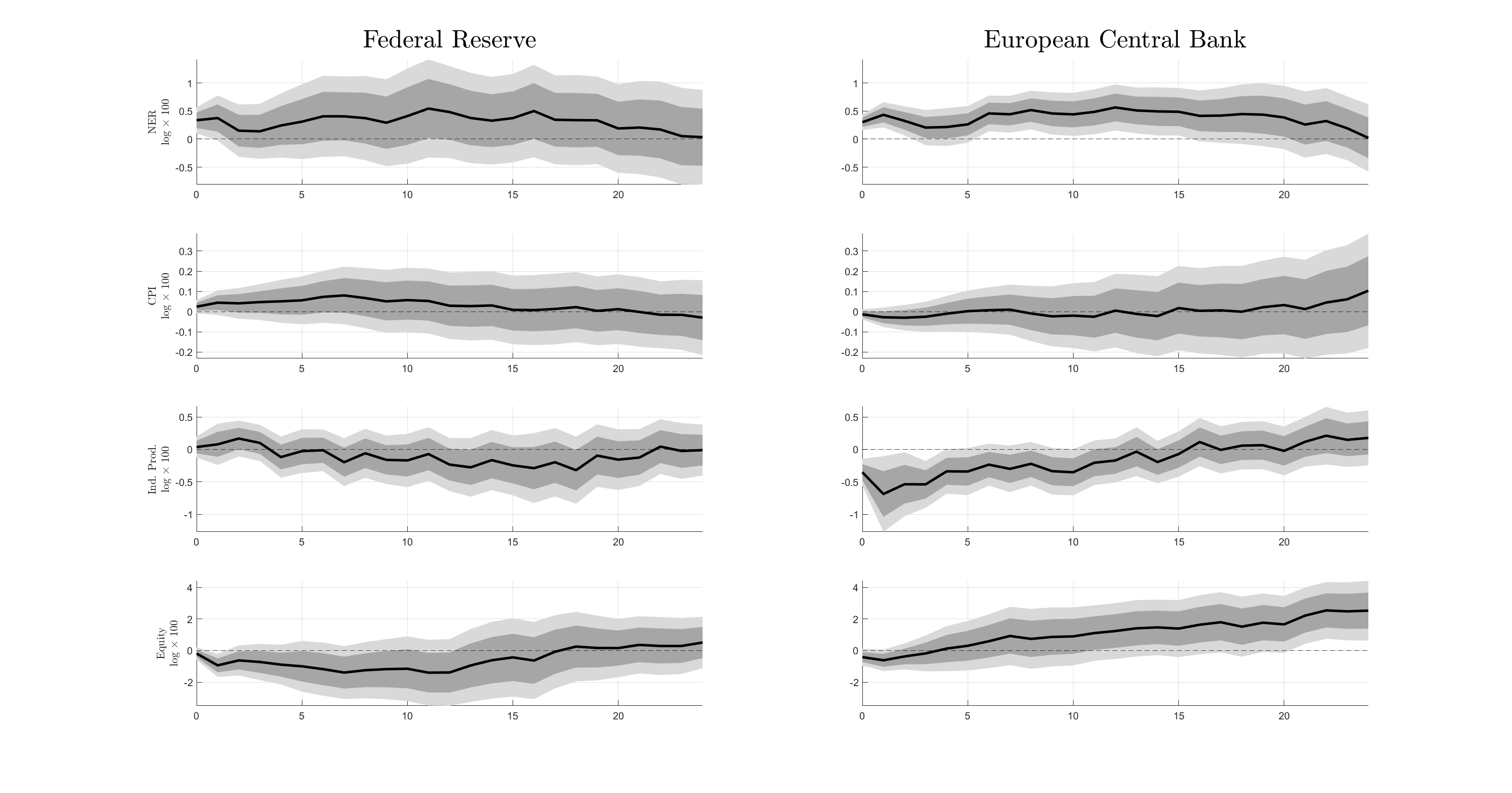}
    \caption{Linear International Spillovers of Monetary Policy}
    \label{fig:Linear_Spillovers_Fed_vs_ECB}
    \floatfoot{\textbf{Note:} This figure reports impulse response functions estimated from linear (symmetric) local projections to monetary policy shocks originating in the United States (left column) and the euro area (right column).
    The four rows report responses of the nominal exchange rate (NER), consumer prices (CPI), industrial production, and equity prices.
    The nominal exchange rate is expressed as units of domestic currency per U.S. dollar (left column) and per euro (right column), so an increase denotes a depreciation of the domestic currency.
    Solid lines denote point estimates. Dark shaded areas indicate 68\% confidence intervals, while light shaded areas indicate 90\% confidence intervals.}
\end{figure}

\noindent
\textbf{Linear spillovers.}
Figure \ref{fig:Linear_Spillovers_Fed_vs_ECB} reports impulse responses estimated under a linear (symmetric) local projection specification for Federal Reserve and European Central Bank monetary policy shocks. On average, monetary policy tightenings in both jurisdictions are associated with economically meaningful international spillovers. Foreign currencies depreciate, consumer prices rise, industrial production declines, and equity prices fall following a contractionary monetary policy shock.

These responses are broadly consistent with the existing literature on international monetary spillovers. However, the linear specification imposes symmetry between expansionary and contractionary shocks and therefore implicitly assumes that monetary policy easings undo the effects of tightenings. The remainder of this section shows that this assumption is strongly rejected by the data.

\medskip

\medskip

\begin{landscape}
\begin{figure}
    \centering
    \includegraphics[width=14cm,height=10cm]{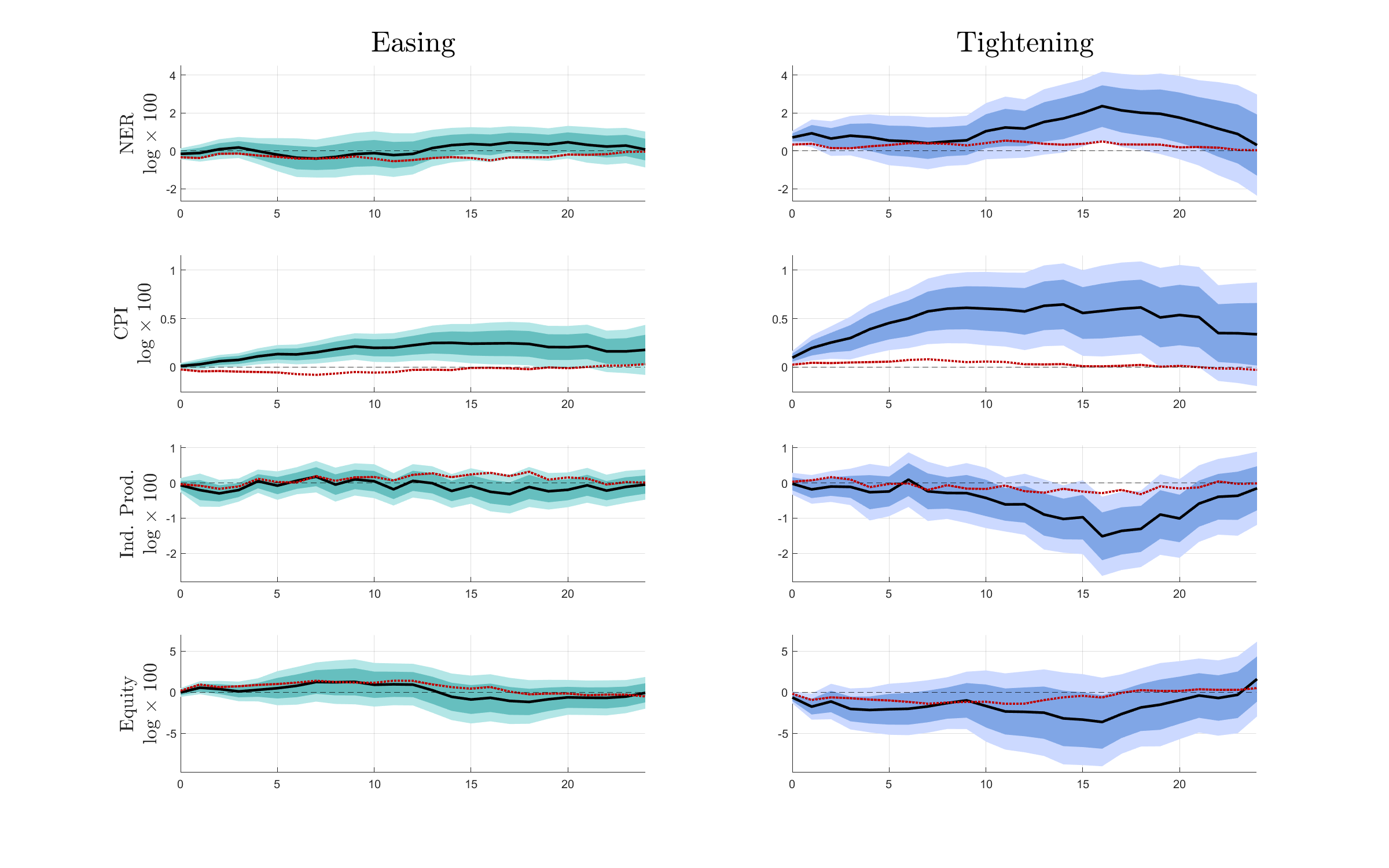}
    \caption{Sign-Dependent International Spillovers of U.S. Monetary Policy}
    \label{fig:FED_Asymmetry_AbsSign}
    \floatfoot{\textbf{Note:} This figure reports impulse response functions to U.S. monetary policy shocks estimated using sign-dependent local projections.
    The figure consists of twelve panels arranged in four rows and three columns.
    Rows correspond to (from top to bottom) the nominal exchange rate (NER), consumer prices (CPI), industrial production, and equity prices.
    The left column reports responses from a linear (symmetric) specification.
    The middle column reports responses to expansionary (easing) monetary policy shocks, while the right column reports responses to contractionary (tightening) monetary policy shocks.
    The nominal exchange rate is expressed as units of domestic currency per U.S. dollar, so an increase denotes a depreciation of the domestic currency.
    Solid lines denote point estimates. Dark shaded areas indicate 68\% confidence intervals, while light shaded areas indicate 90\% confidence intervals.}
\end{figure}
\end{landscape}

\noindent
\textbf{Federal Reserve.}
Figure \ref{fig:FED_Asymmetry_AbsSign} reports sign-dependent impulse responses to U.S. monetary policy shocks. Allowing for sign dependence reveals substantial nonlinearities in the international transmission of Federal Reserve policy.

Contractionary U.S. monetary policy shocks generate significantly larger and more persistent spillovers than expansionary shocks of comparable magnitude. Following a Fed tightening, foreign currencies experience a pronounced nominal depreciation against the U.S. dollar, with effects that peak at medium horizons and remain elevated for several years. In contrast, Fed easings do not generate a statistically significant exchange rate response at any horizon, with estimated effects remaining close to zero throughout.

The asymmetry is particularly pronounced for prices and real activity. A Fed tightening leads to a much larger and statistically significant increase in foreign consumer prices, consistent with strong exchange rate pass-through during contractionary episodes. Fed easings are also associated with an increase in the price index, but the magnitude of this response is substantially smaller. As a result, the linear specification, which averages over easings and tightenings, fails to detect a statistically significant price response.

Real activity exhibits a similar pattern. A U.S. monetary tightening produces a large and persistent contraction in foreign industrial production, with the trough occurring roughly one to one-and-a-half years after impact. These effects are economically meaningful and statistically significant. By contrast, expansionary shocks do not generate any detectable increase in industrial production at any horizon, with estimated responses remaining close to zero. This asymmetry explains why the linear specification yields muted average effects on output.

Financial spillovers also differ by sign. Foreign equity prices fall following both easings and tightenings, but the decline associated with a Fed tightening is slightly larger and more persistent. In contrast, equity price responses to easings are smaller in magnitude. The linear specification therefore masks meaningful differences in the transmission of contractionary and expansionary U.S. monetary policy shocks.

\medskip

\begin{landscape}
\begin{figure}
    \centering
    \includegraphics[width=14cm,height=10cm]{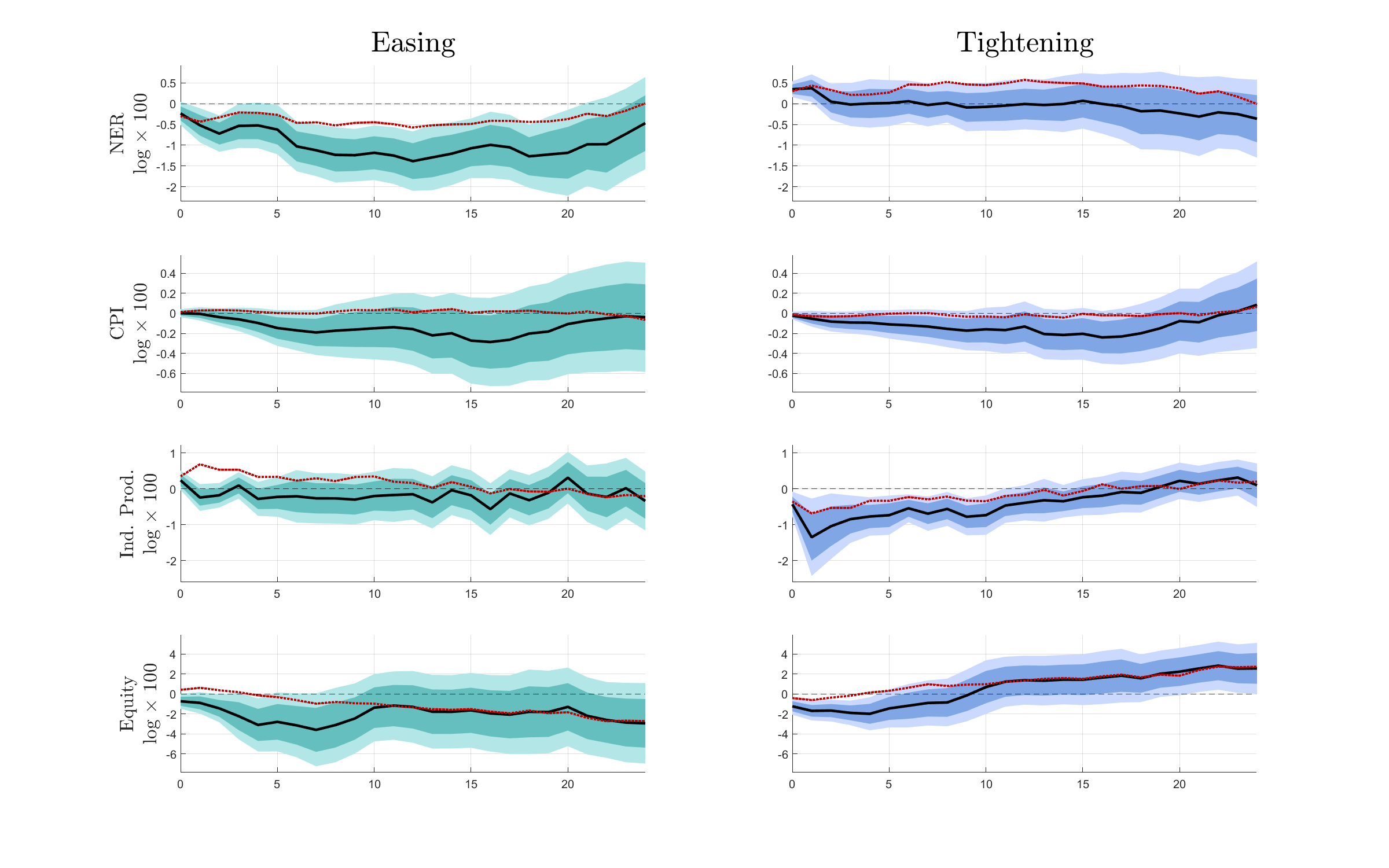}
    \caption{Sign-Dependent International Spillovers of Euro Area Monetary Policy}
    \label{fig:ECB_Symmetric_vs_Asymmetric}
    \floatfoot{\textbf{Note:} This figure reports impulse response functions to euro area monetary policy shocks estimated using sign-dependent local projections.
    The figure consists of twelve panels arranged in four rows and three columns.
    Rows correspond to (from top to bottom) the nominal exchange rate (NER), consumer prices (CPI), industrial production, and equity prices.
    The left column reports responses from a linear (symmetric) specification.
    The middle column reports responses to expansionary (easing) monetary policy shocks, while the right column reports responses to contractionary (tightening) monetary policy shocks.
    The nominal exchange rate is expressed as units of domestic currency per euro, so an increase denotes a depreciation of the domestic currency.
    Solid lines denote point estimates. Dark shaded areas indicate 68\% confidence intervals, while light shaded areas indicate 90\% confidence intervals.}
\end{figure}
\end{landscape}

\noindent
\textbf{European Central Bank.}
Figure \ref{fig:ECB_Symmetric_vs_Asymmetric} reports the corresponding results for euro area monetary policy shocks. As in the U.S. case, the linear specification suggests economically meaningful international spillovers on average. Allowing for sign dependence, however, reveals an even starker asymmetry.

International spillovers from ECB monetary policy are overwhelmingly driven by contractionary shocks. Following an ECB tightening, foreign currencies depreciate against the euro, although the exchange rate response is somewhat smaller than in the U.S. case. Despite this more moderate exchange rate movement, the effects on prices, real activity, and financial markets are substantially larger.

ECB tightenings lead to a much larger decline in the foreign price index, a pronounced and persistent contraction in industrial production, and a substantial fall in equity prices. These responses are economically large and statistically significant across a wide range of horizons. In contrast, ECB easings generate a stronger exchange rate appreciation than predicted by the linear specification, but this appreciation does not translate into improved real or financial outcomes. Industrial production does not increase following an ECB easing, and equity prices experience a significant decline rather than an expansion.

As a result, the linear specification substantially misrepresents the international transmission of euro area monetary policy by averaging strong contractionary effects with weak or adverse responses to expansionary shocks.

\medskip

\noindent
\textbf{Summary.}
Across both central banks, international spillovers are strongly sign-dependent. Contractionary monetary policy shocks generate large and statistically significant spillovers to foreign prices, real activity, and financial markets, while expansionary shocks do not produce a symmetric reversal. For the Federal Reserve, output and exchange rate responses to easings remain close to zero, and price responses are substantially smaller than those following tightenings. For the ECB, spillovers are even more asymmetric, with contractionary shocks accounting for virtually all economically meaningful international effects. These findings imply that linear specifications substantially understate the international transmission of contractionary monetary policy shocks.

%%%%%%%%%%%%%%%%%%%%%%%%%%%%%%%%%%%%%%%%%%%%%%%%%%%%%%%%%%%%%%%%%%%%%%
%%%%%%%%%%%%%%%%%%%%%%%%%%%%%%%%%%%%%%%%%%%%%%%%%%%%%%%%%%%%%%%%%%%%%%
\section{Understanding the Asymmetry in Spillovers}
\label{sec:understanding_asymmetry}

The results in Section \ref{sec:sign_dependent_spillovers} show that international monetary spillovers are strongly sign-dependent and that linear specifications substantially misrepresent both their magnitude and, in some cases, their direction. In this section, we examine the transmission channels underlying these asymmetries. Our focus is on how monetary policy shocks propagate internationally through financial conditions and trade, and how these channels differ systematically across contractionary and expansionary episodes.

We proceed in three steps. First, we document how international financial conditions respond asymmetrically to monetary policy shocks of different signs, focusing on interest rates, risk premia, and cross-border capital flows. Second, we examine the implications of these financial responses for international trade, showing that trade volumes and prices adjust primarily during episodes of monetary tightening. Finally, we address the puzzling international effects of euro area monetary policy easings and argue that they can be understood through policy interactions and global financial dominance, whereby the international transmission of ECB policy is shaped by the endogenous response of U.S. monetary policy and global financial conditions.

%%%%%%%%%%%%%%%%%%%%%%%%%%%%%%%%%%%%%%%%%%%%%%%%%%%%%%%%%%%%%%%%%%%%%%
\subsection{Financial Conditions in International Spillovers}
\label{subsec:financial_conditions}

To shed light on the mechanisms underlying sign-dependent spillovers, this subsection examines how key price-based and quantity-based measures of international transmission respond to monetary policy shocks of different signs. We focus on interest rates and sovereign spreads, which capture the cost of external finance, as well as cross-border capital flows, which reflect the quantity of international financial intermediation. For each central bank, we present sign-dependent impulse responses separately for rates and for flows, allowing for a transparent comparison between expansionary and contractionary shocks.

\begin{figure}
    \centering
    \includegraphics[width=14cm,height=10cm]{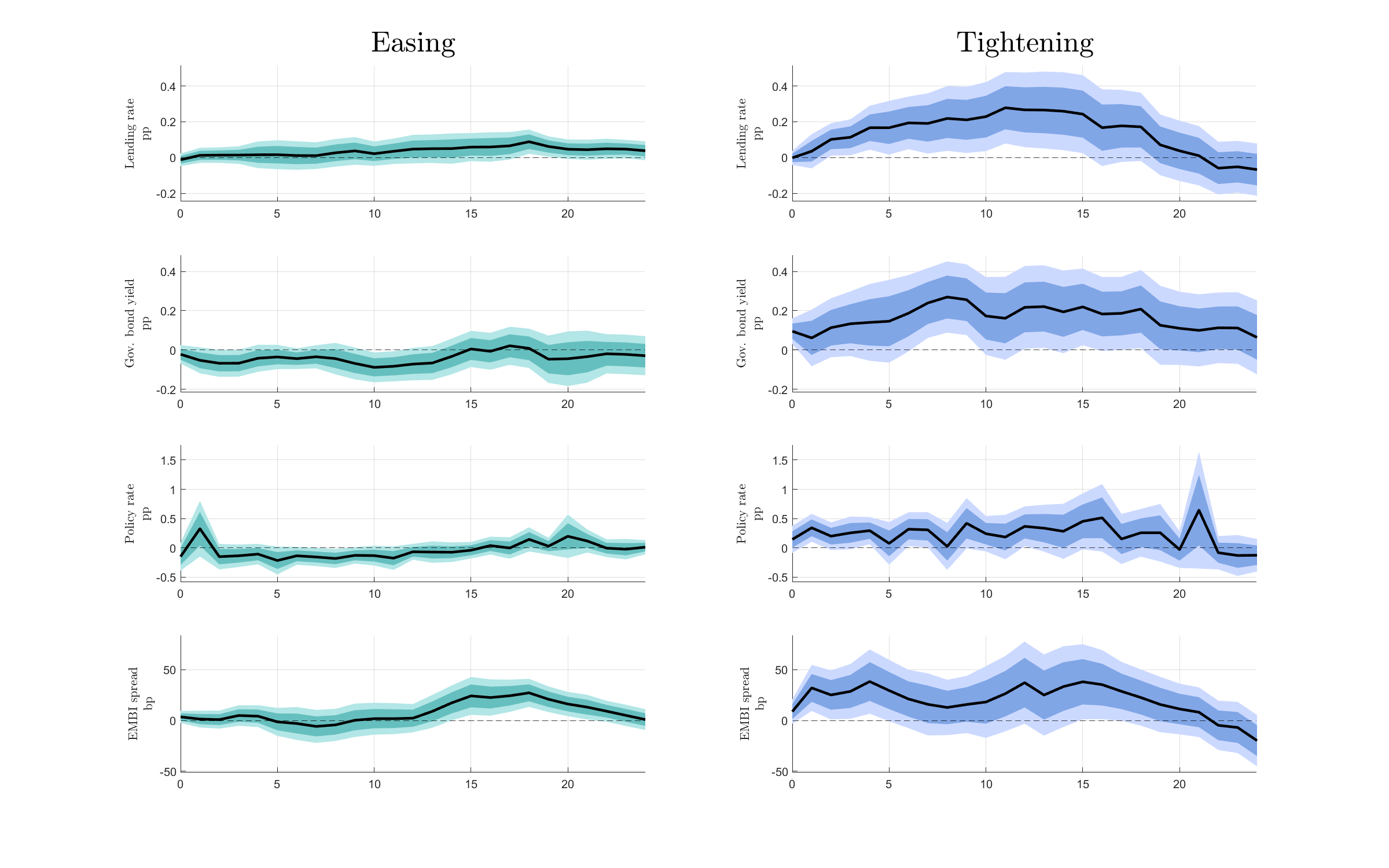}
    \caption{Sign-Dependent Responses of International Rates to U.S. Monetary Policy}
    \label{fig:FED_Rates_Asymmetry}
    \floatfoot{
    \textbf{Note:} This figure reports sign-dependent impulse response functions of international interest rates and sovereign spreads to U.S. monetary policy shocks.
    The left column shows responses to expansionary (easing) shocks, while the right column shows responses to contractionary (tightening) shocks.
    Rows correspond to lending rates, government bond yields, policy rates, and EMBI spreads.
    Solid lines denote point estimates.
    Dark shaded areas indicate 68\% confidence intervals, while light shaded areas indicate 90\% confidence intervals.
    Responses are reported over a 24-month horizon.
    }
\end{figure}

\begin{figure}
    \centering
    \includegraphics[width=14cm,height=10cm]{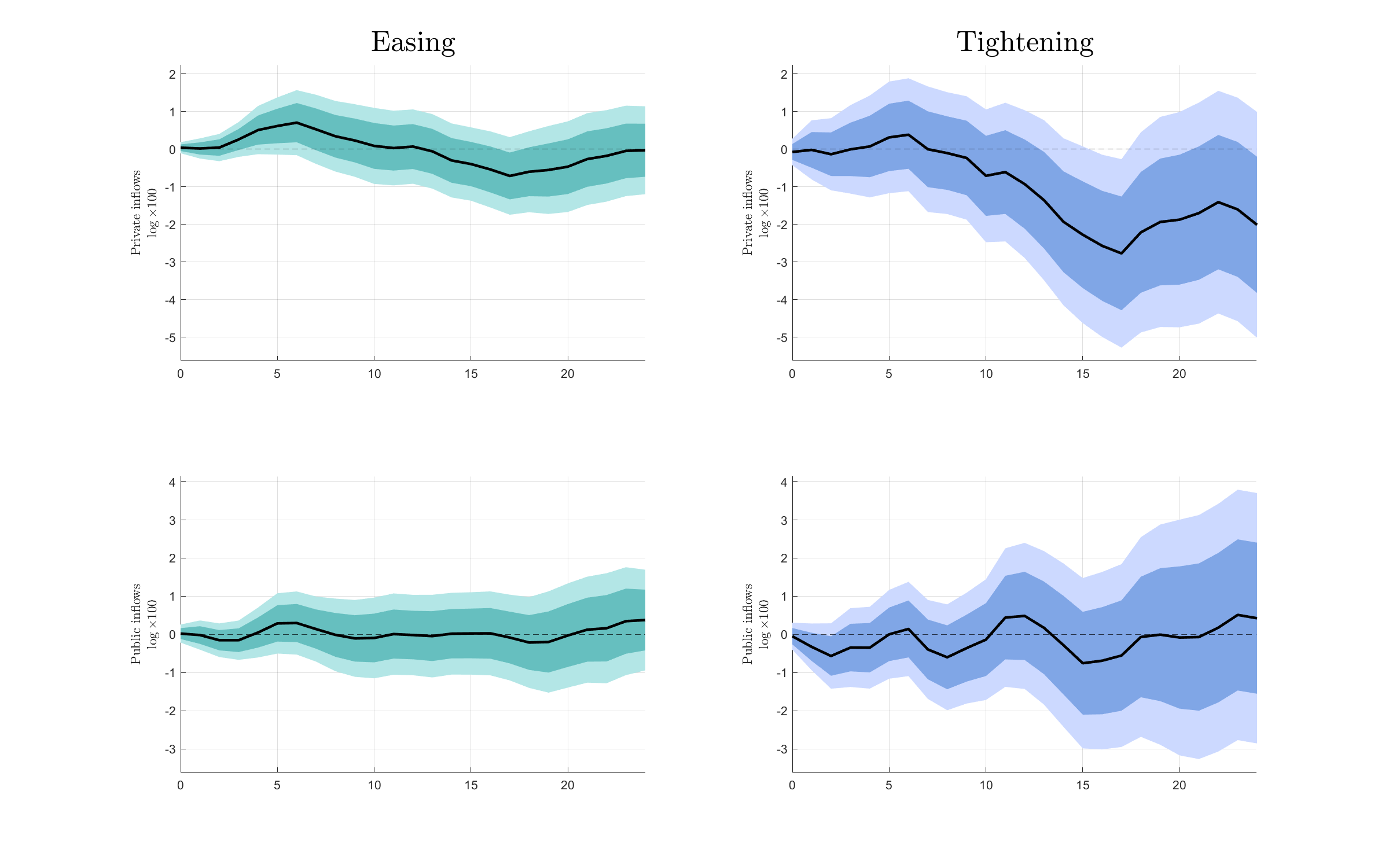}
    \caption{Sign-Dependent Responses of Cross-Border Capital Flows to U.S. Monetary Policy}
    \label{fig:FED_Flows_Asymmetry}
    \floatfoot{
    \textbf{Note:} This figure reports sign-dependent impulse response functions of cross-border capital flows to U.S. monetary policy shocks.
    The left column shows responses to expansionary (easing) shocks, while the right column shows responses to contractionary (tightening) shocks.
    Rows correspond to private and public capital inflows, expressed as log changes multiplied by 100.
    Solid lines denote point estimates.
    Dark shaded areas indicate 68\% confidence intervals, while light shaded areas indicate 90\% confidence intervals.
    Responses are reported over a 24-month horizon.
    }
\end{figure}

\noindent
\textbf{Federal Reserve.}
Figures \ref{fig:FED_Rates_Asymmetry} and \ref{fig:FED_Flows_Asymmetry} report sign-dependent impulse responses of international interest rates, sovereign spreads, and cross-border capital flows to U.S. monetary policy shocks.

Contractionary U.S. monetary policy shocks generate a clear and persistent tightening of international financial conditions. Following a Fed tightening, foreign lending rates and government bond yields increase significantly, and EMBI spreads widen sharply. These effects are economically large and remain elevated over medium horizons, indicating a sustained increase in the cost of external finance faced by foreign economies.

By contrast, expansionary U.S. monetary policy shocks generate little improvement in borrowing conditions. Lending rates and government bond yields decline only modestly, while EMBI spreads exhibit limited compression. Across several horizons, estimated responses remain close to zero and statistically insignificant. This asymmetry mirrors the macroeconomic evidence documented in Section \ref{sec:sign_dependent_spillovers}, where Federal Reserve easings fail to generate meaningful expansions in output or equity prices abroad.

Cross-border capital flows respond in a similarly asymmetric manner. Federal Reserve tightenings trigger a pronounced and persistent contraction in private capital inflows, with effects that peak at medium horizons and remain economically meaningful thereafter. Public capital inflows also decline following tightenings, though their response is more muted. In contrast, U.S. monetary policy easings do not generate a sustained expansion in international capital flows. Private inflows rise only modestly on impact and quickly revert, while public inflows display little systematic response throughout the horizon.

Taken together, these results indicate that contractionary U.S. monetary policy shocks propagate internationally through both price-based and quantity-based financial channels, while expansionary shocks do not symmetrically reverse these effects.

\begin{figure}
    \centering
    \includegraphics[width=14cm,height=10cm]{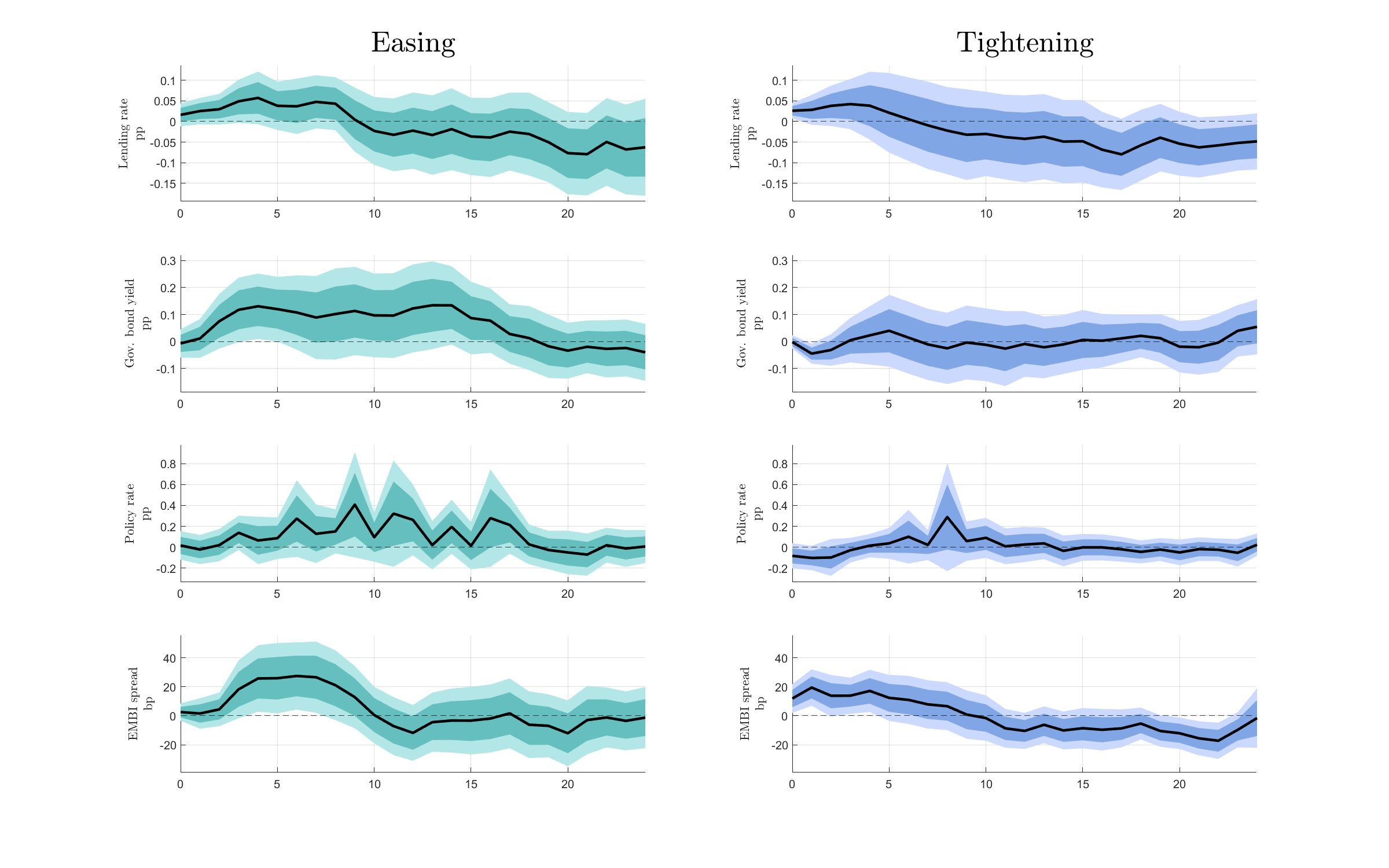}
    \caption{Euro Area Monetary Policy and International Interest Rates}
    \label{fig:ECB_Financial_Conditions_Rates}
    \floatfoot{
    \textbf{Note:} This figure reports sign-dependent impulse response functions of international interest rates and sovereign spreads to euro area monetary policy shocks.
    The left column shows responses to expansionary (easing) shocks, while the right column shows responses to contractionary (tightening) shocks.
    Rows correspond to lending rates, government bond yields, policy rates, and EMBI spreads.
    Solid lines denote point estimates.
    Dark shaded areas indicate 68\% confidence intervals, while light shaded areas indicate 90\% confidence intervals.
    Responses are reported over a 24-month horizon.
    }
\end{figure}

\begin{figure}
    \centering
    \includegraphics[width=14cm,height=10cm]{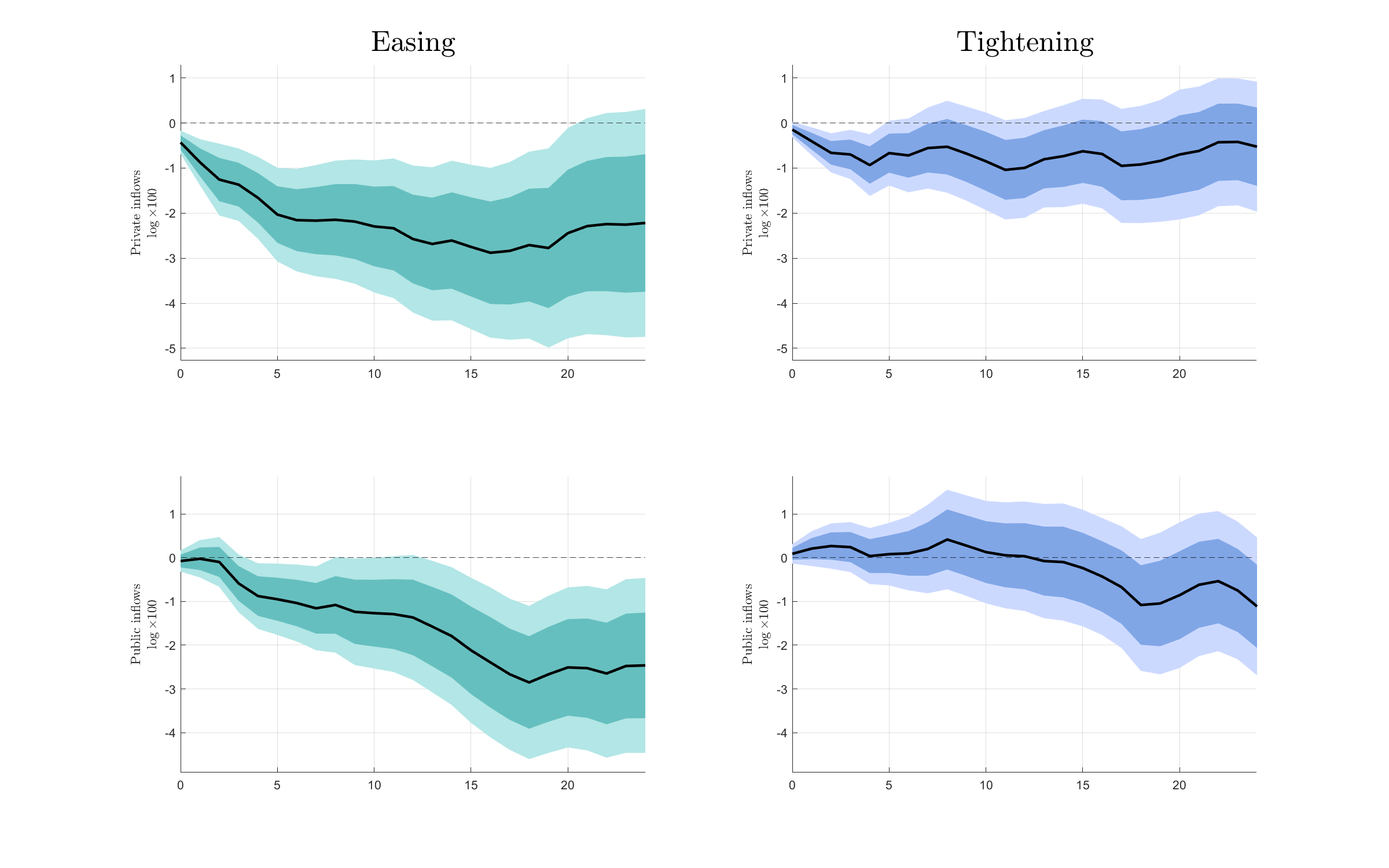}
    \caption{Euro Area Monetary Policy and Cross-Border Capital Flows}
    \label{fig:ECB_Financial_Conditions_Flows}
    \floatfoot{
    \textbf{Note:} This figure reports sign-dependent impulse response functions of cross-border capital flows to euro area monetary policy shocks.
    The left column shows responses to expansionary (easing) shocks, while the right column shows responses to contractionary (tightening) shocks.
    The top row reports private capital inflows, while the bottom row reports public capital inflows.
    Capital flows are expressed as log changes multiplied by 100.
    Solid lines denote point estimates.
    Dark shaded areas indicate 68\% confidence intervals, while light shaded areas indicate 90\% confidence intervals.
    Responses are reported over a 24-month horizon.
    }
\end{figure}

\medskip
\noindent
\textbf{European Central Bank.}
Figures \ref{fig:ECB_Financial_Conditions_Rates} and \ref{fig:ECB_Financial_Conditions_Flows} report the corresponding sign-dependent responses to euro area monetary policy shocks.

As in the U.S. case, contractionary euro area monetary policy shocks are associated with a marked tightening of international financial conditions. Following an ECB tightening, lending rates, government bond yields, and policy rates increase persistently, while sovereign risk spreads widen substantially. These responses are economically large and statistically significant across a wide range of horizons.

Cross-border capital flows reinforce this pattern. Private capital inflows decline sharply following ECB tightenings, with effects that deepen over time and remain persistent at medium horizons. Public inflows also contract, though more gradually. In contrast, expansionary ECB monetary policy shocks do not generate a symmetric improvement in financial conditions. Interest rates and sovereign spreads decline only modestly, and capital inflows fail to increase. As a result, expansionary euro area monetary policy shocks are associated with a deterioration, rather than an improvement, in several dimensions of international financial conditions.

%%%%%%%%%%%%%%%%%%%%%%%%%%%%%%%%%%%%%%%%%%%%%%%%%%%%%%%%%%%%%%%%%%%%%%
\subsection{Trade in International Spillovers}
\label{subsec:trade_conditions}

We next examine the implications of sign-dependent monetary policy shocks for international trade. Trade flows provide a complementary perspective on international transmission, as they capture the response of real cross-border activity rather than financial prices or quantities alone. If the asymmetries documented in financial conditions are economically meaningful, they should also be reflected in the behavior of exports, imports, and trade prices.

\bigskip

\begin{figure}
    \centering
    \includegraphics[width=14cm,height=10cm]{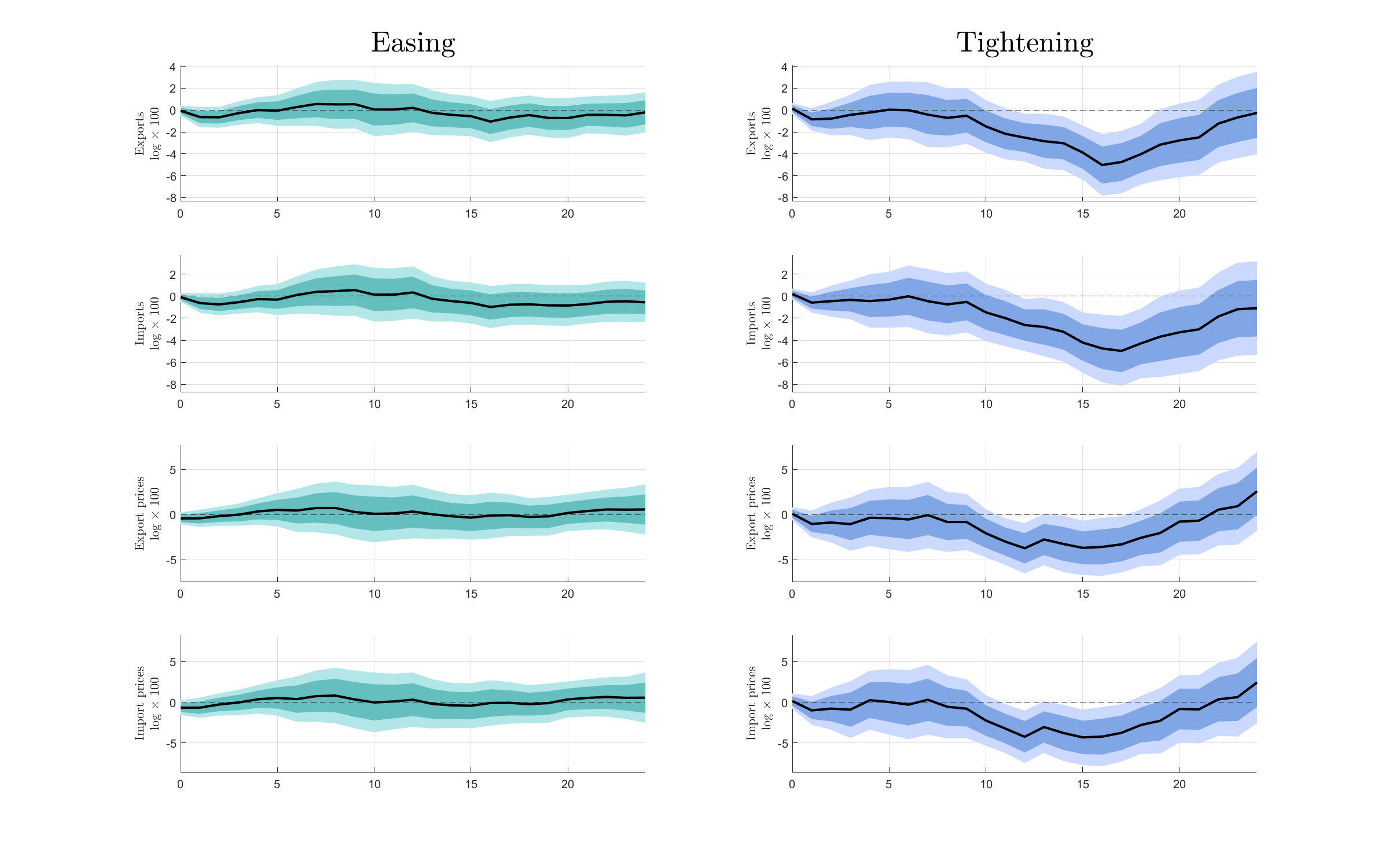}
    \caption{International Trade Responses to U.S. Monetary Policy}
    \label{fig:FED_Trade_Easing_vs_Tightening}
    \floatfoot{
    \textbf{Note:} This figure reports sign-dependent impulse response functions of international trade volumes and trade prices to U.S. monetary policy shocks.
    The left column shows responses to expansionary (easing) shocks, while the right column shows responses to contractionary (tightening) shocks.
    From top to bottom, the panels report responses of exports, imports, export prices, and import prices.
    All variables are expressed as log changes multiplied by 100.
    Solid lines denote point estimates.
    Dark shaded areas indicate 68\% confidence intervals, while light shaded areas indicate 90\% confidence intervals.
    Responses are reported over a 24-month horizon.
    }
\end{figure}

\noindent
\textbf{Federal Reserve.}
Figure \ref{fig:FED_Trade_Easing_vs_Tightening} reports sign-dependent impulse responses of international trade volumes and trade prices to U.S. monetary policy shocks. Contractionary U.S. monetary policy shocks generate large and persistent declines in international trade. Exports and imports fall sharply following a Fed tightening, and trade prices decline in tandem. By contrast, expansionary U.S. monetary policy shocks do not generate a symmetric improvement in trade outcomes. Trade volumes and prices remain close to zero throughout the horizon, implying that linear specifications substantially understate the contractionary impact of U.S. monetary policy on international trade.

\begin{figure}
    \centering
    \includegraphics[width=14cm,height=10cm]{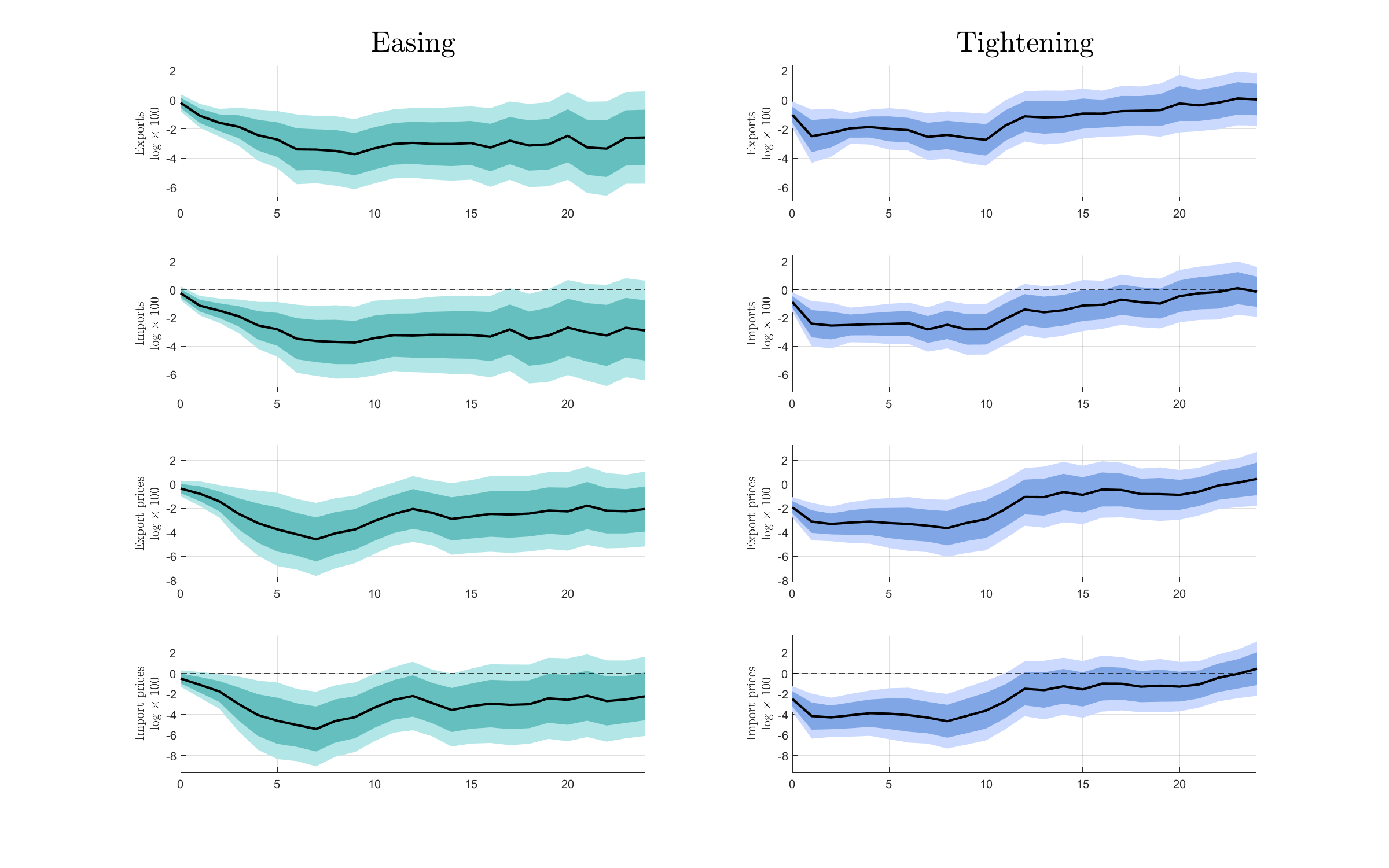}
    \caption{International Trade Responses to Euro Area Monetary Policy}
    \label{fig:ECB_Trade_Easing_vs_Tightening}
    \floatfoot{
    \textbf{Note:} This figure reports sign-dependent impulse response functions of international trade volumes and trade prices to euro area monetary policy shocks.
    The left column shows responses to expansionary (easing) shocks, while the right column shows responses to contractionary (tightening) shocks.
    From top to bottom, the panels report responses of exports, imports, export prices, and import prices.
    All variables are expressed as log changes multiplied by 100.
    Solid lines denote point estimates.
    Dark shaded areas indicate 68\% confidence intervals, while light shaded areas indicate 90\% confidence intervals.
    Responses are reported over a 24-month horizon.
    }
\end{figure}

\medskip
\noindent
\textbf{European Central Bank.}
Figure \ref{fig:ECB_Trade_Easing_vs_Tightening} reports the corresponding trade responses to euro area monetary policy shocks. As in the U.S. case, contractionary euro area monetary policy shocks lead to a pronounced contraction in international trade. Strikingly, expansionary ECB monetary policy shocks also generate a deterioration in trade volumes and prices. As a result, both expansionary and contractionary euro area monetary policy shocks are associated with a decline in international trade, highlighting a puzzle that cannot be reconciled with standard linear models of international monetary transmission.

%%%%%%%%%%%%%%%%%%%%%%%%%%%%%%%%%%%%%%%%%%%%%%%%%%%%%%%%%%%%%%%%%%%%%
\subsection{Why Do ECB Easings Fail to Generate International Expansions?}
\label{subsec:ecb_puzzle}

The previous sections show that contractionary monetary policy shocks originating in both the United States and the euro area are associated with a significant tightening of global financial conditions and sharp declines in international trade and commodity prices. At the same time, the international effects of expansionary monetary policy differ markedly across central banks. While Federal Reserve easings do not generate a statistically significant improvement in global financial conditions or trade, euro area monetary policy easings are followed by a pronounced worsening of both. Following an ECB easing, financial conditions tighten globally, trade volumes contract, and commodity prices fall sharply.

Ex ante, these outcomes are surprising, as expansionary monetary policy is typically associated with improved financial conditions and stronger economic activity. In this subsection, we shed light on this puzzle by examining the response of U.S. monetary policy and dollar-denominated global financial conditions to euro area monetary policy shocks. We show that ECB easings are followed by an endogenous tightening of U.S. monetary policy, a persistent appreciation of the U.S. dollar, and a broad deterioration in global financial and trade conditions, which helps reconcile the contractionary international effects of expansionary euro area monetary policy.

\bigskip

\begin{figure}
    \centering
    \includegraphics[width=14cm,height=10cm]{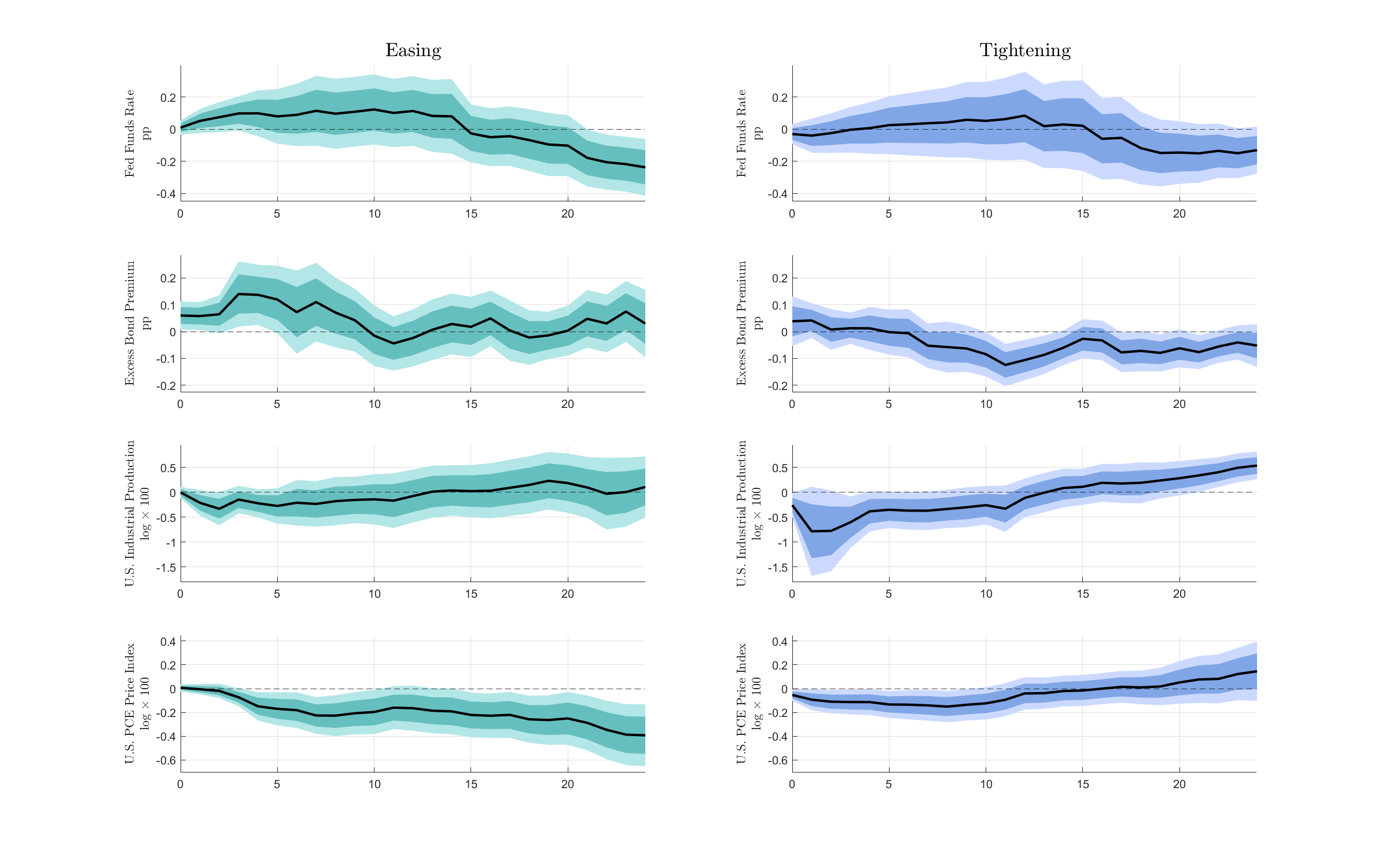}
    \caption{U.S. Responses to Euro Area Monetary Policy Shocks}
    \label{fig:ECB_US_Response}
    \floatfoot{
    \textbf{Note:} This figure reports sign-dependent impulse response functions of U.S. variables to euro area monetary policy shocks.
    Rows correspond to the federal funds rate, the excess bond premium (EBP), U.S. industrial production, and the U.S. price index.
    The left column shows responses to expansionary (easing) euro area monetary policy shocks, while the right column shows responses to contractionary (tightening) shocks.
    Impulse responses are estimated using the benchmark specification, including euro area controls.
    Solid lines denote point estimates.
    Dark shaded areas indicate 68\% confidence intervals, while light shaded areas indicate 90\% confidence intervals.
    Standard errors are clustered at the date level.
    Responses are reported over a 24-month horizon.
    }
\end{figure}

Figure \ref{fig:ECB_US_Response} reports sign-dependent impulse responses of key U.S. macroeconomic and financial variables to euro area monetary policy shocks. The figure shows responses of the federal funds rate, the excess bond premium (EBP), U.S. industrial production, and the U.S. price index, estimated using the benchmark specification with euro area controls and standard errors clustered at the date level.

Following an ECB tightening, U.S. industrial production and the U.S. price index decline significantly, indicating negative spillovers into the U.S. real economy. Financial conditions worsen modestly, with an initial increase in the excess bond premium, although this response is not precisely estimated. Importantly, ECB tightenings do not elicit a systematic response from U.S. monetary policy: the federal funds rate remains close to zero throughout the horizon. Overall, these responses are consistent with ECB tightenings being associated with a global downturn that spills over to the United States, without triggering an endogenous tightening by the Federal Reserve.

In contrast, the responses to ECB easings are strikingly different. Following an expansionary euro area monetary policy shock, the federal funds rate increases persistently and significantly, indicating an endogenous tightening of U.S. monetary policy. This tightening is accompanied by a deterioration in U.S. financial conditions, as reflected in a rise in the excess bond premium. At the same time, U.S. industrial production contracts and the U.S. price index falls. Thus, while ECB tightenings are associated with negative U.S. spillovers in the absence of a Federal Reserve response, ECB easings are followed by contractionary U.S. outcomes precisely because they trigger an endogenous tightening of U.S. monetary policy.

\bigskip

\begin{figure}
    \centering
    \includegraphics[width=14cm,height=10cm]{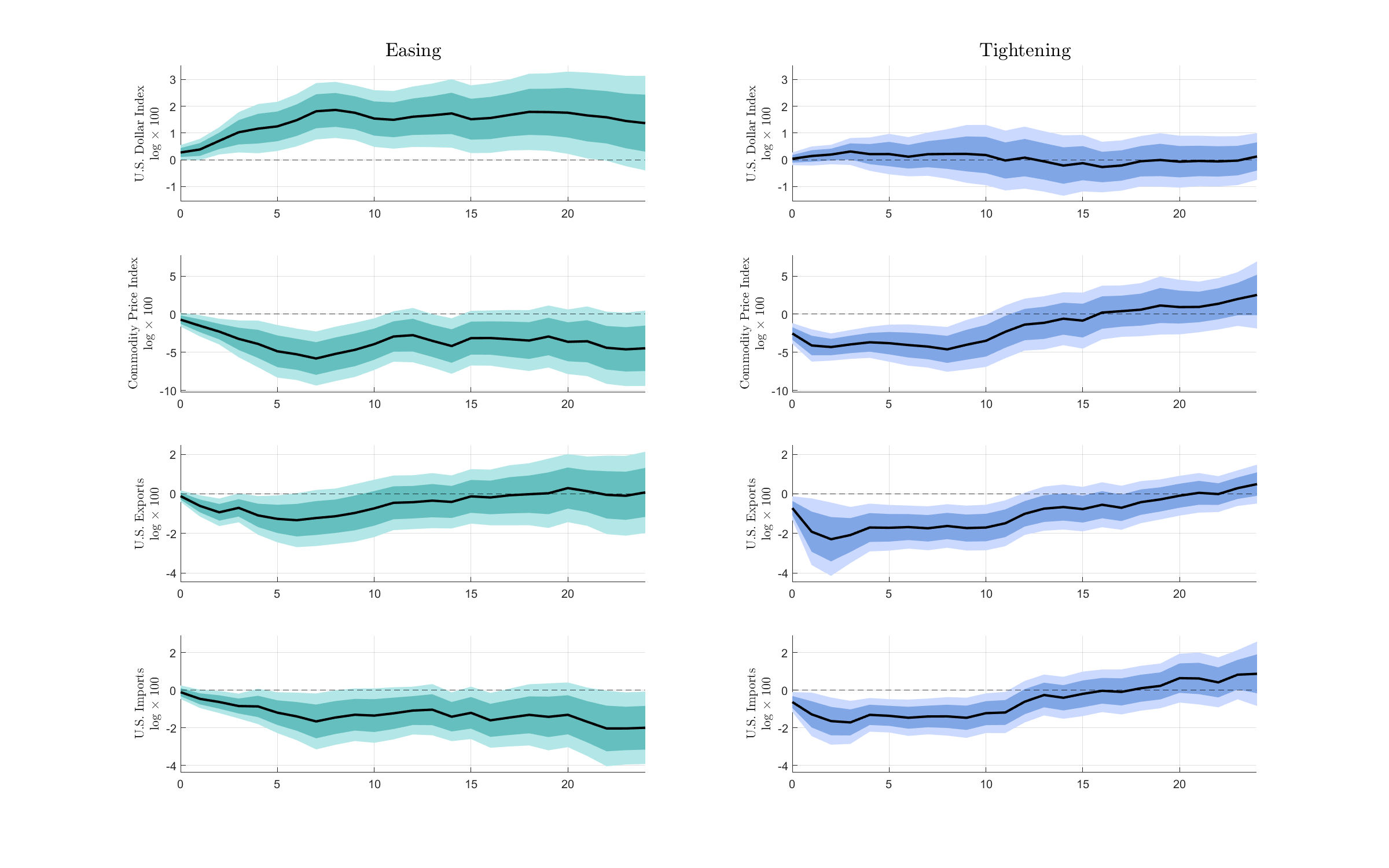}
    \caption{Dollar, Commodity Prices, and U.S. Trade Responses to Euro Area Monetary Policy}
    \label{fig:ECB_US_Dollar_Trade}
    \floatfoot{
    \textbf{Note:} This figure reports sign-dependent impulse response functions of the U.S. dollar index, a world commodity price index, U.S. exports, and U.S. imports to euro area monetary policy shocks.
    Each variable is added one at a time to the benchmark specification, while maintaining the same set of euro area controls.
    The left column shows responses to expansionary (easing) euro area monetary policy shocks, while the right column shows responses to contractionary (tightening) shocks.
    All variables are expressed as log changes multiplied by 100.
    Solid lines denote point estimates.
    Dark shaded areas indicate 68\% confidence intervals, while light shaded areas indicate 90\% confidence intervals.
    Standard errors are clustered at the date level.
    Responses are reported over a 24-month horizon.
    }
\end{figure}

Figure \ref{fig:ECB_US_Dollar_Trade} extends the analysis by examining the response of the U.S. dollar, global commodity prices, and U.S. trade to euro area monetary policy shocks. The figure reports sign-dependent impulse responses of the U.S. dollar index, a world commodity price index, U.S. exports, and U.S. imports. Each variable is added one at a time to the benchmark specification, while maintaining the same euro area controls and clustering standard errors at the date level.

Consistent with the limited response of U.S. monetary policy, ECB tightenings do not generate a significant appreciation or depreciation of the U.S. dollar. Nevertheless, ECB tightenings are followed by a pronounced decline in world commodity prices and by sizable contractions in both U.S. exports and U.S. imports. These responses indicate that ECB tightenings are associated with a global recessionary episode that depresses commodity demand and international trade, spilling over to the United States through real rather than exchange-rate channels.

By contrast, ECB easings are associated with a markedly different pattern. Following an expansionary euro area monetary policy shock, the U.S. dollar appreciates strongly and persistently. This dollar appreciation coincides with a sharp decline in world commodity prices and with large and persistent reductions in U.S. exports and imports. Taken together with the responses documented in Figure \ref{fig:ECB_US_Response}, these results indicate that ECB easings are followed by a tightening of dollar-denominated global financial conditions driven by an endogenous Federal Reserve response. This tightening is closely associated with the contraction in commodity prices and international trade observed after ECB easings.

%%%%%%%%%%%%%%%%%%%%%%%%%%%%%%%%%%%%%%%%%%%%%%%%%%%%%%%%%%%%%%%%%%%%%%
%%%%%%%%%%%%%%%%%%%%%%%%%%%%%%%%%%%%%%%%%%%%%%%%%%%%%%%%%%%%%%%%%%%%%%
\section{Additional Results \& Robustness Checks}
\label{sec:additional_results_robustness}

This section evaluates the robustness of the sign-dependent international monetary spillovers documented in Section~\ref{sec:sign_dependent_spillovers}. We assess robustness along four dimensions. First, we consider alternative high-frequency identification strategies for monetary policy shocks. Second, we evaluate alternative parameterizations of sign dependence that allow for greater flexibility in the response to easing and tightening shocks. Third, we examine sensitivity to key specification choices, including lag length and the inclusion of fixed effects. Finally, we assess robustness across subsamples defined by country groups and time periods, including Advanced versus Non-Advanced Economies, the pre-COVID period, and the post–Global Financial Crisis (GFC) era. Across all exercises, the central finding of asymmetric international spillovers—driven primarily by tightening episodes—remains unchanged.

%%%%%%%%%%%%%%%%%%%%%%%%%%%%%%%%%%%%%%%%%%%%
\subsection{Identification Robustness}
\label{subsec:additional_identification}

\paragraph{Bauer--Swanson identification.}
We first assess the robustness of our main results to an alternative identification of U.S. monetary policy shocks. While the benchmark analysis relies on the high-frequency shocks constructed by \citet{jarocinski2020deconstructing}, here we instead use the orthogonalized monetary policy shock proposed by \citet{bauer2022reassessment}, which is designed to purge information effects and isolate exogenous monetary policy innovations.

The empirical specification is otherwise unchanged. We estimate both linear (symmetric) and sign-dependent local projections using the absolute--sign decomposition, maintain the same set of controls and sample period, and cluster standard errors at the date level. This exercise therefore isolates the role of shock identification while holding all other elements of the estimation fixed.

\begin{figure}
    \centering
    \includegraphics[width=14cm,height=10cm]{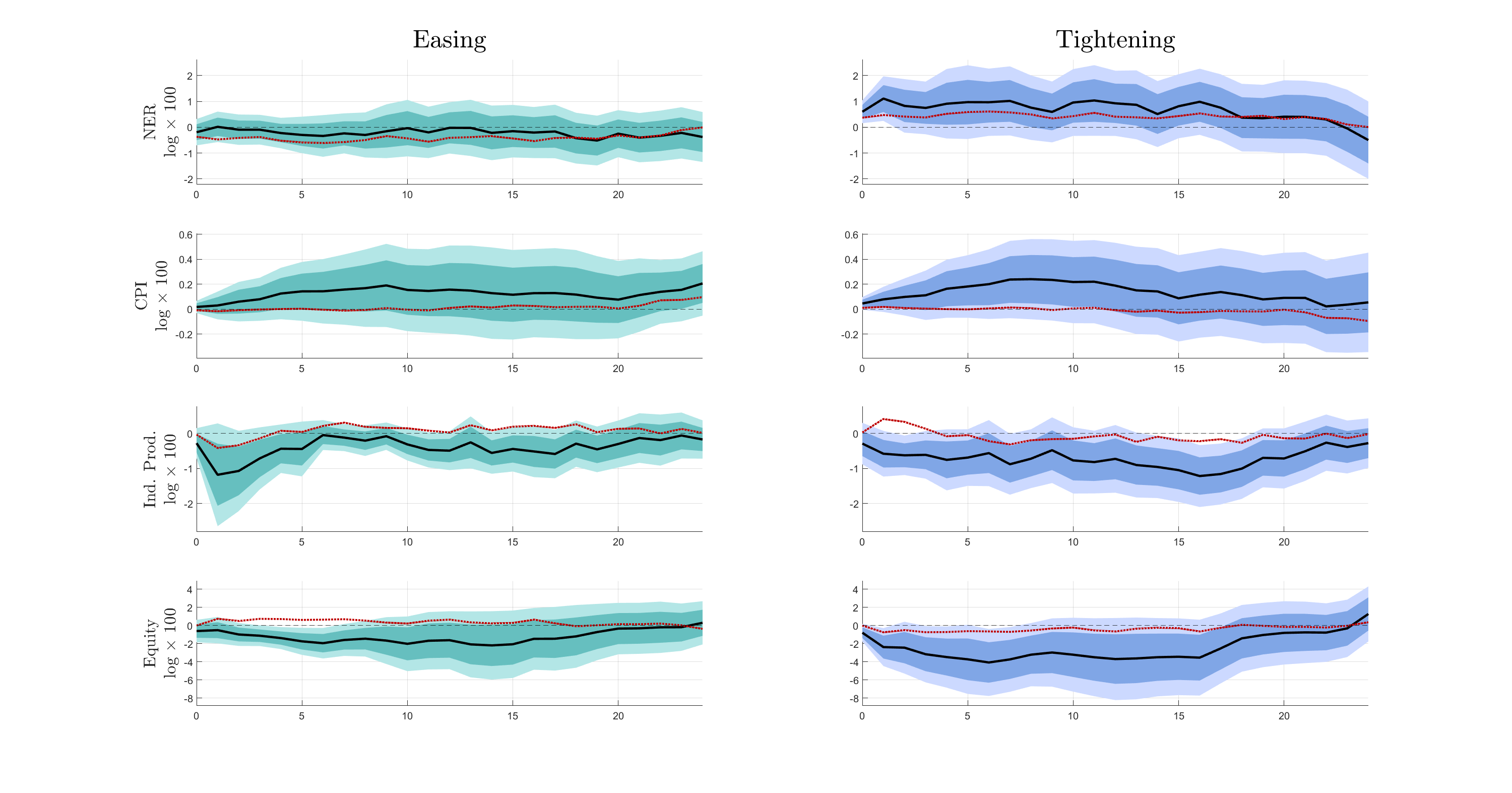}
    \caption{Sign-Dependent International Spillovers of U.S. Monetary Policy: Bauer--Swanson Shocks}
    \label{fig:FED_BS_Asymmetry}
    \floatfoot{
    \textbf{Note:} This figure reports impulse response functions to U.S. monetary policy shocks identified using the orthogonalized shock of \citet{bauer2022reassessment}. Rows correspond to the nominal exchange rate (NER), consumer prices (CPI), industrial production, and equity prices. The left column reports responses to expansionary (easing) shocks, while the right column reports responses to contractionary (tightening) shocks. Solid lines denote point estimates. Dark shaded areas indicate 68\% confidence intervals, while light shaded areas indicate 90\% confidence intervals. Red dotted lines report the linear (symmetric) benchmark impulse responses, sign-flipped in the easing column. All specifications mirror the benchmark estimation and include the same set of controls. Standard errors are clustered at the date level. Responses are reported over a 24-month horizon.
    }
\end{figure}

Figure~\ref{fig:FED_BS_Asymmetry} reports the resulting impulse responses. The results closely mirror those obtained in the benchmark analysis. Contractionary U.S. monetary policy shocks generate large and persistent international spillovers, including a depreciation of foreign currencies against the U.S. dollar, higher consumer prices, a pronounced contraction in real activity, and a sustained decline in equity prices. By contrast, expansionary shocks produce substantially weaker responses across all outcomes.

Allowing for sign dependence again reveals a pronounced asymmetry in international transmission. Linear specifications understate the magnitude of contractionary spillovers by averaging over responses to easings and tightenings that differ sharply in both magnitude and persistence.

\paragraph{Poor Man’s sign restriction (Federal Reserve).}
We next consider an alternative implementation of the Jarociński--Karadi identification based on the Poor Man’s sign restriction. Instead of relying on the median rotation of high-frequency surprises, this approach classifies monetary policy shocks according to the joint sign of interest rate and stock market surprises, providing a simpler and more conservative separation of monetary policy and information effects.

The empirical specification remains unchanged relative to both the benchmark and the Bauer--Swanson robustness exercise. We estimate sign-dependent local projections using the absolute--sign decomposition, maintain the same controls and sample period, and cluster standard errors at the date level.

\begin{figure}
    \centering
    \includegraphics[width=14cm,height=10cm]{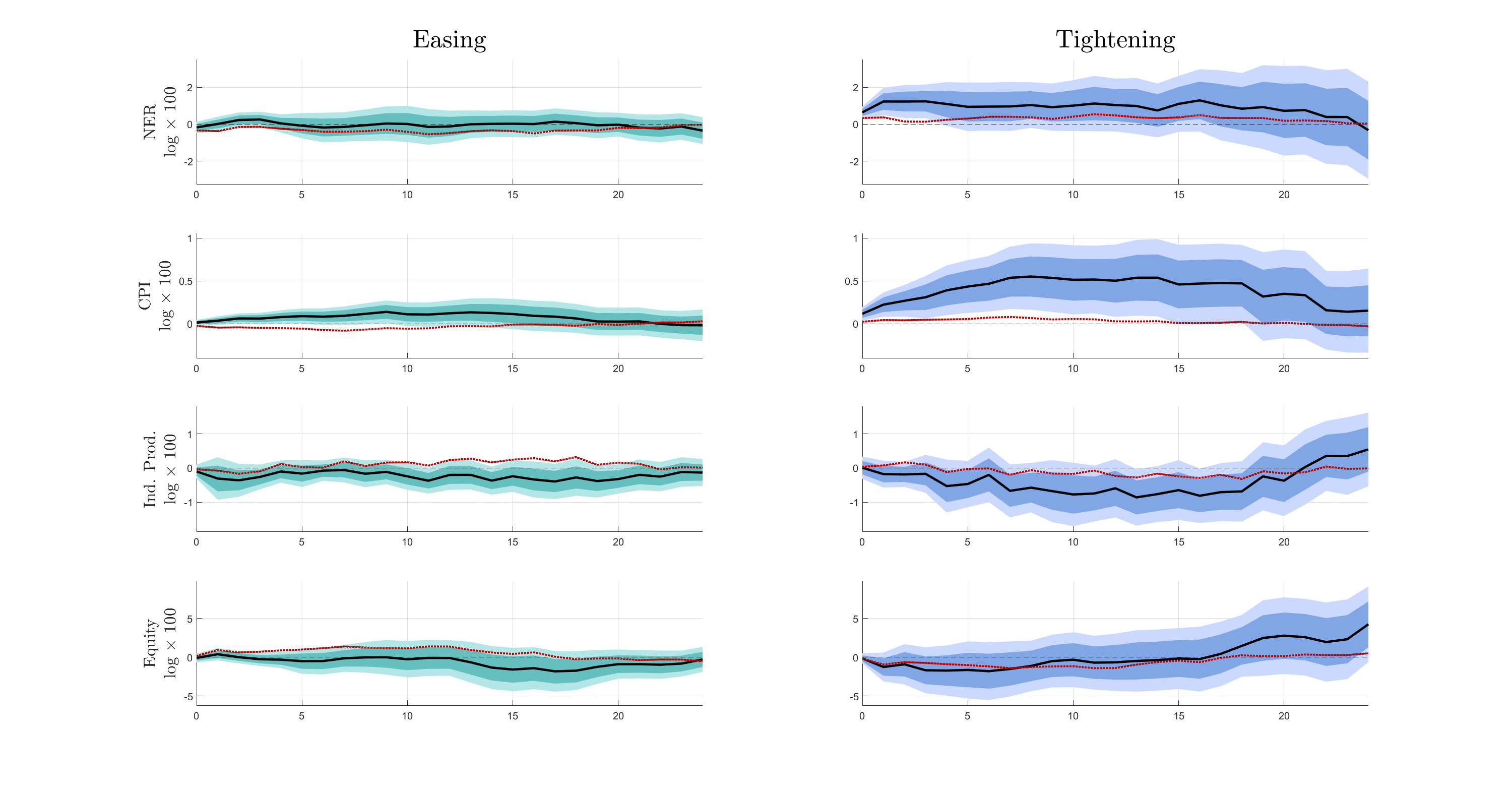}
    \caption{Sign-Dependent International Spillovers of U.S. Monetary Policy: Poor Man’s Sign Restriction}
    \label{fig:FED_PM_Asymmetry}
    \floatfoot{
    \textbf{Note:} This figure reports impulse response functions to U.S. monetary policy shocks identified using the Poor Man’s sign restriction. Rows correspond to the nominal exchange rate (NER), consumer prices (CPI), industrial production, and equity prices. Shaded areas denote 68\% and 90\% confidence intervals. The red dotted line reports the symmetric benchmark response.
    }
\end{figure}

Figure~\ref{fig:FED_PM_Asymmetry} reports the resulting impulse responses. The patterns are again highly consistent with the benchmark results. Contractionary U.S. monetary policy shocks generate sizable and persistent international spillovers across exchange rates, prices, real activity, and equity markets, while expansionary shocks are associated with substantially weaker responses.

\paragraph{Poor Man’s sign restriction (ECB).}
We next assess the robustness of the euro area results to the Poor Man’s sign restriction. The empirical specification is otherwise identical to the benchmark ECB analysis, including the same euro area controls, sample period, and inference procedure.

\begin{figure}
    \centering
    \includegraphics[width=14cm,height=10cm]{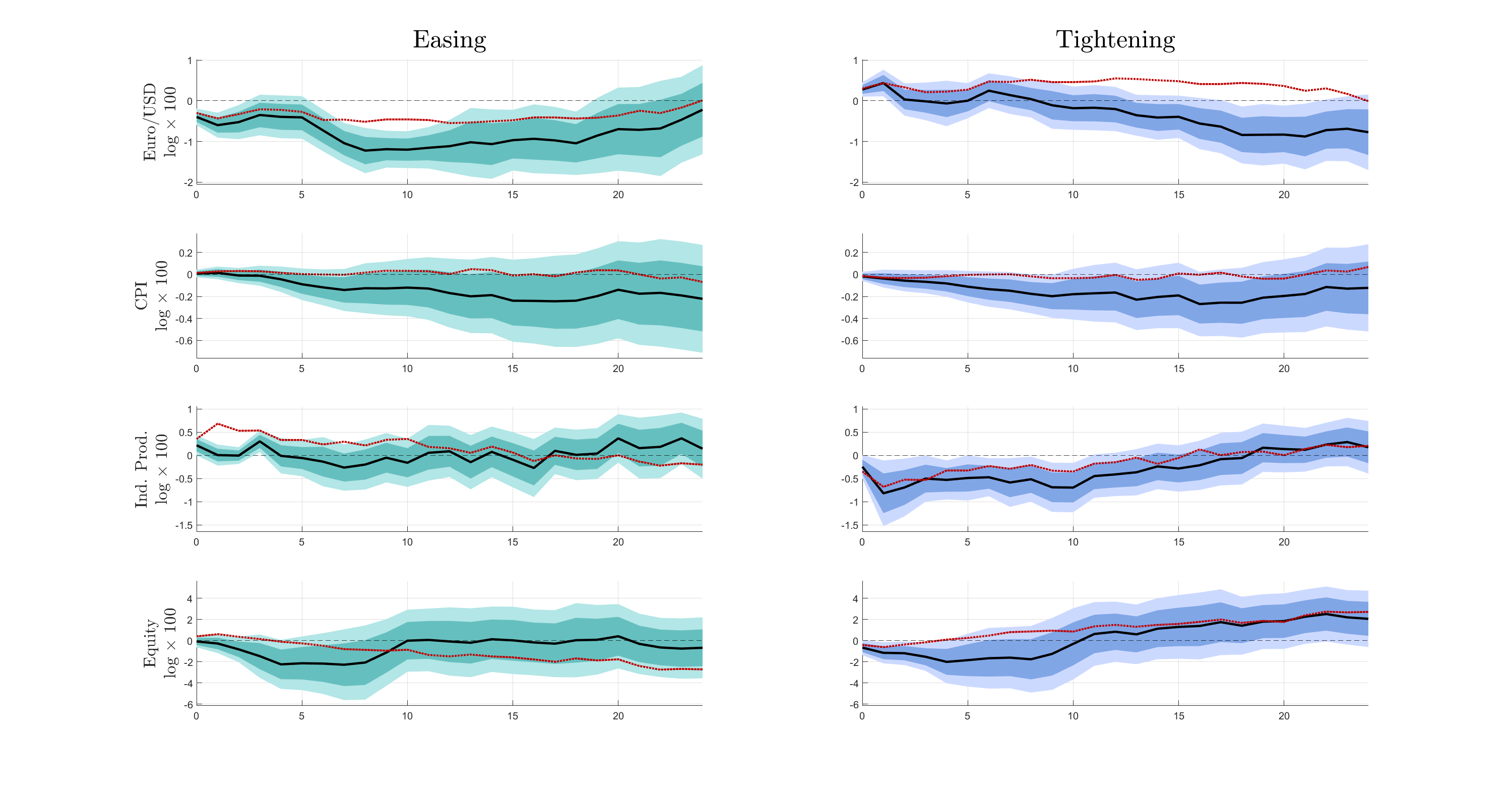}
    \caption{Sign-Dependent International Spillovers of Euro Area Monetary Policy: Poor Man’s Sign Restriction}
    \label{fig:ECB_PM_Asymmetry}
    \floatfoot{
    \textbf{Note:} This figure reports impulse response functions to ECB monetary policy shocks identified using the Poor Man’s sign restriction. Rows correspond to the euro exchange rate, CPI, industrial production, and equity prices. Shaded areas denote 68\% and 90\% confidence intervals.
    }
\end{figure}

Figure~\ref{fig:ECB_PM_Asymmetry} reports the resulting impulse responses. Contractionary ECB shocks generate sizable and persistent international spillovers, including exchange rate movements, declines in consumer prices, contractions in industrial production, and persistent equity price declines. Expansionary ECB shocks again fail to generate a symmetric international expansion.

%%%%%%%%%%%%%%%%%%%%%%%%%%%%%%%%%%%%%%%%%%%%%%%%%%%%%%%%%%%%%%%%%%%%%%
%%%%%%%%%%%%%%%%%%%%%%%%%%%%%%%%%%%%%%%%%%%%%%%%%%%%%%%%%%%%%%%%%%%%%%
\subsection{Specification Robustness: Alternative Treatments of Sign Dependence}
\label{subsec:additional_spec_signrobust}

This subsection examines whether the sign-dependent international spillovers documented in the benchmark analysis are sensitive to the particular functional form used to model asymmetry. The baseline specification adopts an absolute--sign decomposition, which parameterizes impulse responses as a function of the monetary policy shock and its absolute value. Under symmetry of the shock distribution, this formulation provides a parsimonious way to capture sign dependence while preserving statistical power.

We consider two alternative and widely used specifications that allow for greater flexibility in how expansionary and contractionary monetary policy shocks affect foreign outcomes. In both cases, the identifying assumptions, sample, control variables, fixed effects, and inference procedures are identical to those in the benchmark analysis. Differences in estimated responses therefore reflect only the treatment of sign dependence.

\paragraph{Piecewise-linear sign interactions.}
We first estimate a piecewise-linear specification that allows the effects of expansionary and contractionary monetary policy shocks to differ freely. Let $\varepsilon_t$ denote the externally identified monetary policy shock, and define its positive and negative components as
\begin{equation}
\varepsilon_t^{+} \equiv \max(\varepsilon_t,0),
\qquad
\varepsilon_t^{-} \equiv \min(\varepsilon_t,0).
\end{equation}
We then estimate local projections of the form
\begin{equation}
y_{i,t+h}
=
\alpha_{i,h}
+
\beta_h^{+}\varepsilon_t^{+}
+
\beta_h^{-}\varepsilon_t^{-}
+
\Gamma_h X_{i,t-1}
+
u_{i,t+h},
\end{equation}
where $y_{i,t+h}$ denotes the outcome variable of interest and $X_{i,t-1}$ collects the same lagged domestic and global controls as in the benchmark specification. As before, all regressions include country-by-calendar-month fixed effects, and standard errors are clustered by date.

\begin{figure}
\centering
\includegraphics[width=\textwidth]{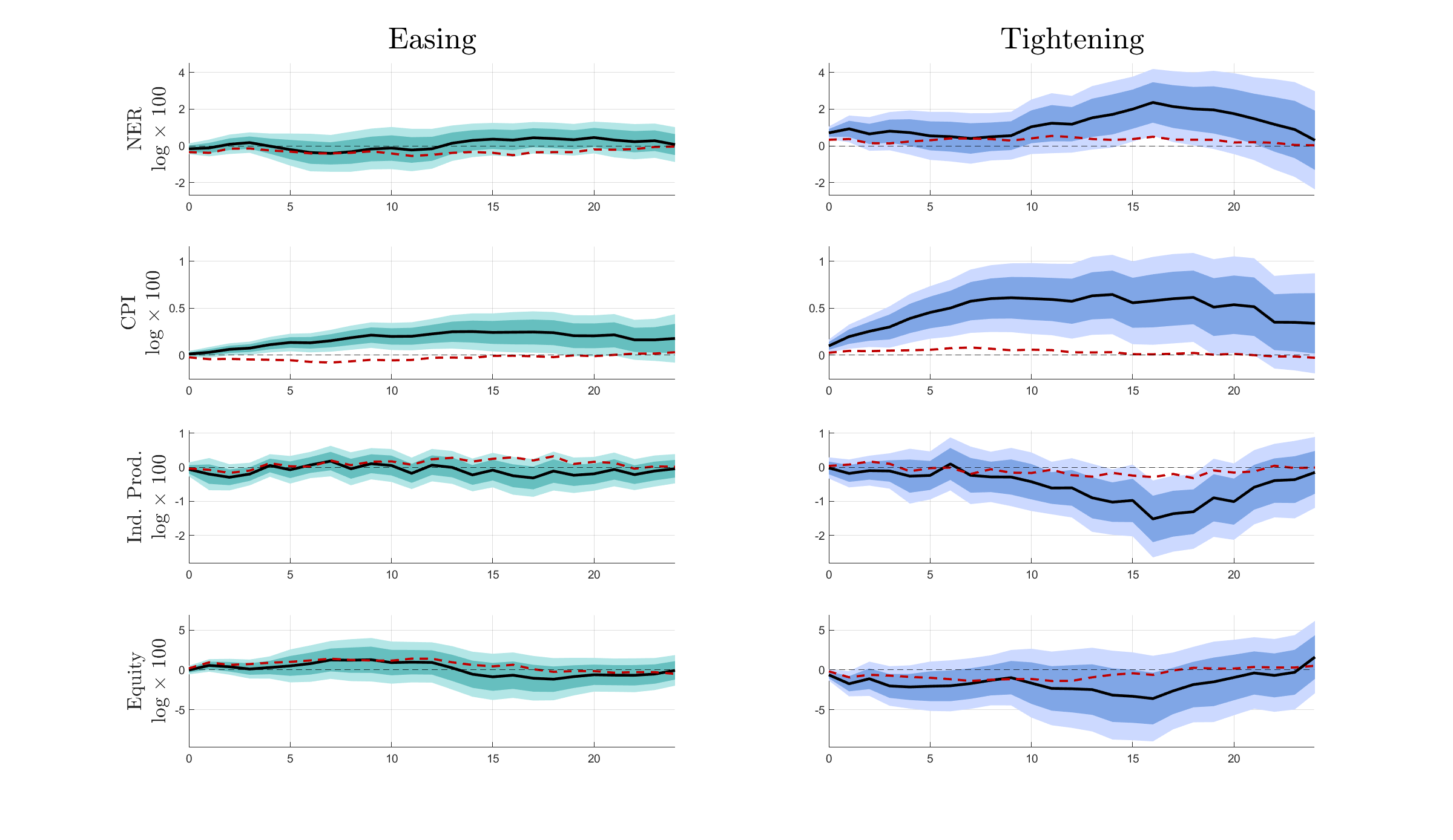}
\caption{Sign-Dependent Spillovers of Federal Reserve Monetary Policy Shocks: Piecewise-Linear Specification.}
\label{fig:fed_signinteraction}
\floatfoot{\footnotesize
\textbf{Notes:} This figure reports impulse responses estimated using a piecewise-linear sign-interaction specification based on $\varepsilon_t^{+}$ and $\varepsilon_t^{-}$. The left column shows responses to easing shocks and the right column shows responses to tightening shocks. Solid black lines denote point estimates. Dark (light) shaded areas correspond to 68\% (90\%) confidence intervals. The red dotted line overlays the symmetric benchmark response (sign-flipped in the easing column). Responses are reported over a 24-month horizon, and standard errors are clustered by date.}
\end{figure}

\begin{figure}
\centering
\includegraphics[width=\textwidth]{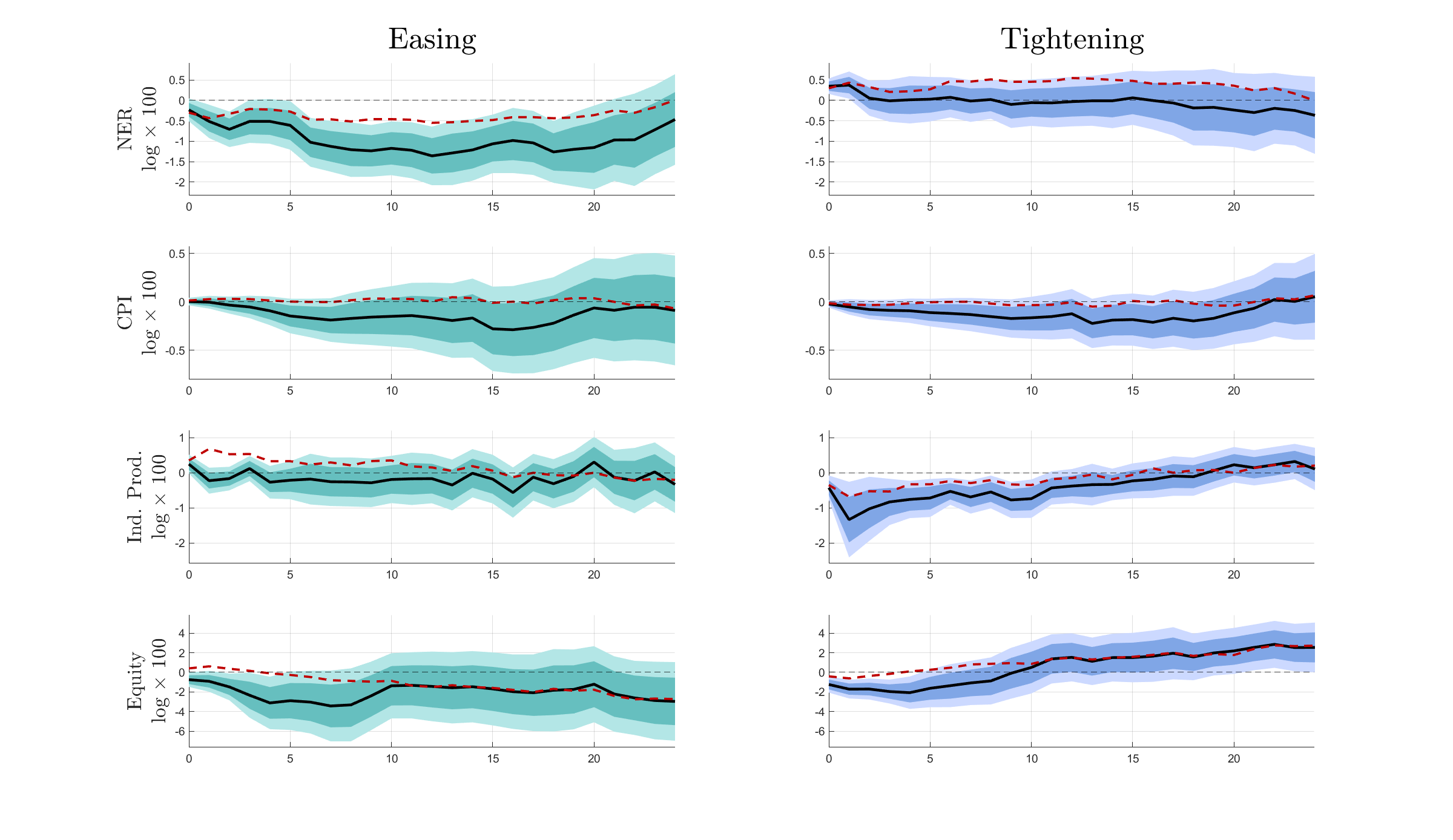}
\caption{Sign-Dependent Spillovers of ECB Monetary Policy Shocks: Piecewise-Linear Specification.}
\label{fig:ecb_signinteraction}
\floatfoot{\footnotesize
\textbf{Notes:} This figure reports impulse responses to ECB monetary policy shocks estimated using the piecewise-linear sign-interaction specification. The left column displays responses to ECB easing shocks and the right column displays responses to ECB tightening shocks. Solid black lines denote point estimates. Dark (light) shaded areas correspond to 68\% (90\%) confidence intervals. The red dotted line overlays the symmetric benchmark response (sign-flipped in the easing column). Responses are reported over a 24-month horizon, and standard errors are clustered by date.}
\end{figure}

Figures~\ref{fig:fed_signinteraction} and~\ref{fig:ecb_signinteraction} report the resulting impulse responses for Federal Reserve and ECB monetary policy shocks, respectively. The piecewise-linear specification yields impulse responses that closely mirror those obtained under the benchmark absolute--sign formulation: tightening shocks generate substantially larger and more persistent international spillovers than easing shocks, while the symmetric benchmark responses lie between the two. These results indicate that the documented asymmetries do not hinge on the absolute--sign parameterization.

\paragraph{Sign-conditioned dynamics with interacted controls.}
We next estimate a more flexible specification in which both the monetary policy shock and the lagged control variables are allowed to have sign-dependent effects, following \citet{ben2023government}. This approach permits the entire dynamic propagation mechanism to differ across tightening and easing episodes, rather than allowing sign dependence only through the contemporaneous shock.

Let
\begin{equation}
D_t \equiv \mathbbm{1}\{\varepsilon_t>0\}
\end{equation}
denote an indicator for tightening shocks. Define sign-conditioned shocks
\begin{equation}
\varepsilon_t^{\mathrm{+}} \equiv \varepsilon_t \cdot D_t,
\qquad
\varepsilon_t^{\mathrm{nonpos}} \equiv \varepsilon_t \cdot (1-D_t),
\end{equation}
and interact the full vector of lagged controls with the same indicator. Concretely, we estimate
\begin{equation}
y_{i,t+h}
=
\alpha_{i,h}
+
\beta_h^{\mathrm{+}}\varepsilon_t^{\mathrm{+}}
+
\beta_h^{\mathrm{-}}\varepsilon_t^{\mathrm{-}}
+
\Gamma_h^{\mathrm{+}}\bigl(D_t \cdot X_{i,t-1}\bigr)
+
\Gamma_h^{\mathrm{-}}\bigl((1-D_t)\cdot X_{i,t-1}\bigr)
+
u_{i,t+h},
\end{equation}
where $X_{i,t-1}$ is the same set of lagged domestic and global controls as in the benchmark specification. All regressions continue to include country-by-calendar-month fixed effects, and standard errors are clustered by date.

\begin{figure}
\centering
\includegraphics[width=\textwidth]{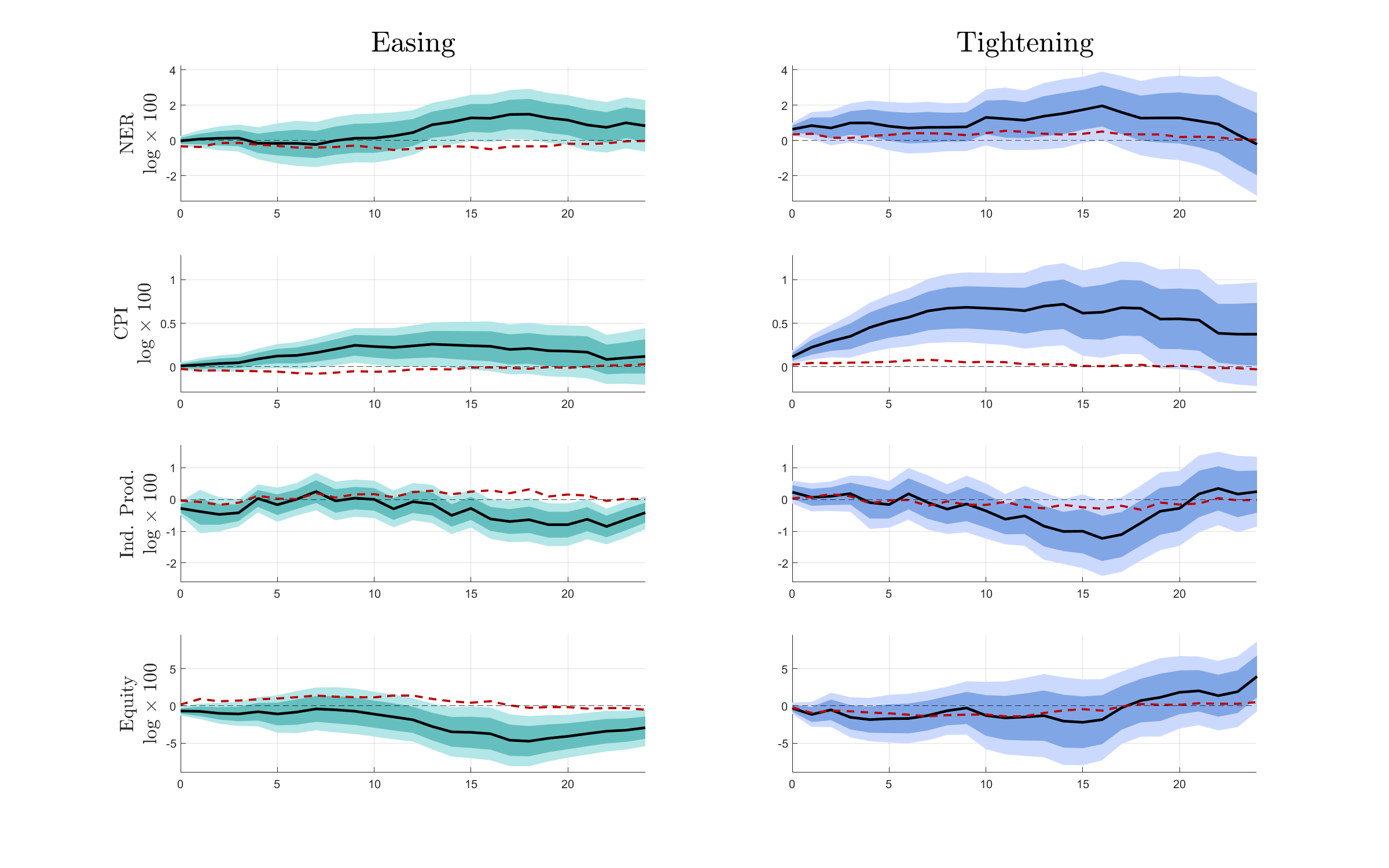}
\caption{Sign-Conditioned Spillovers of Federal Reserve Monetary Policy Shocks: Interacted Controls.}
\label{fig:fed_signconditioned}
\floatfoot{\footnotesize
\textbf{Notes:} This figure reports impulse responses estimated using sign-conditioned local projections in which both the monetary policy shock and the full set of lagged control variables are interacted with the sign indicator $D_t=\mathbbm{1}\{\varepsilon_t>0\}$. The left column shows responses to easing/non-positive shocks and the right column shows responses to tightening shocks. Solid black lines denote point estimates. Dark (light) shaded areas correspond to 68\% (90\%) confidence intervals. The red dotted line overlays the symmetric benchmark response (sign-flipped in the easing column). Responses are reported over a 24-month horizon, and standard errors are clustered by date.}
\end{figure}

\begin{figure}
\centering
\includegraphics[width=\textwidth]{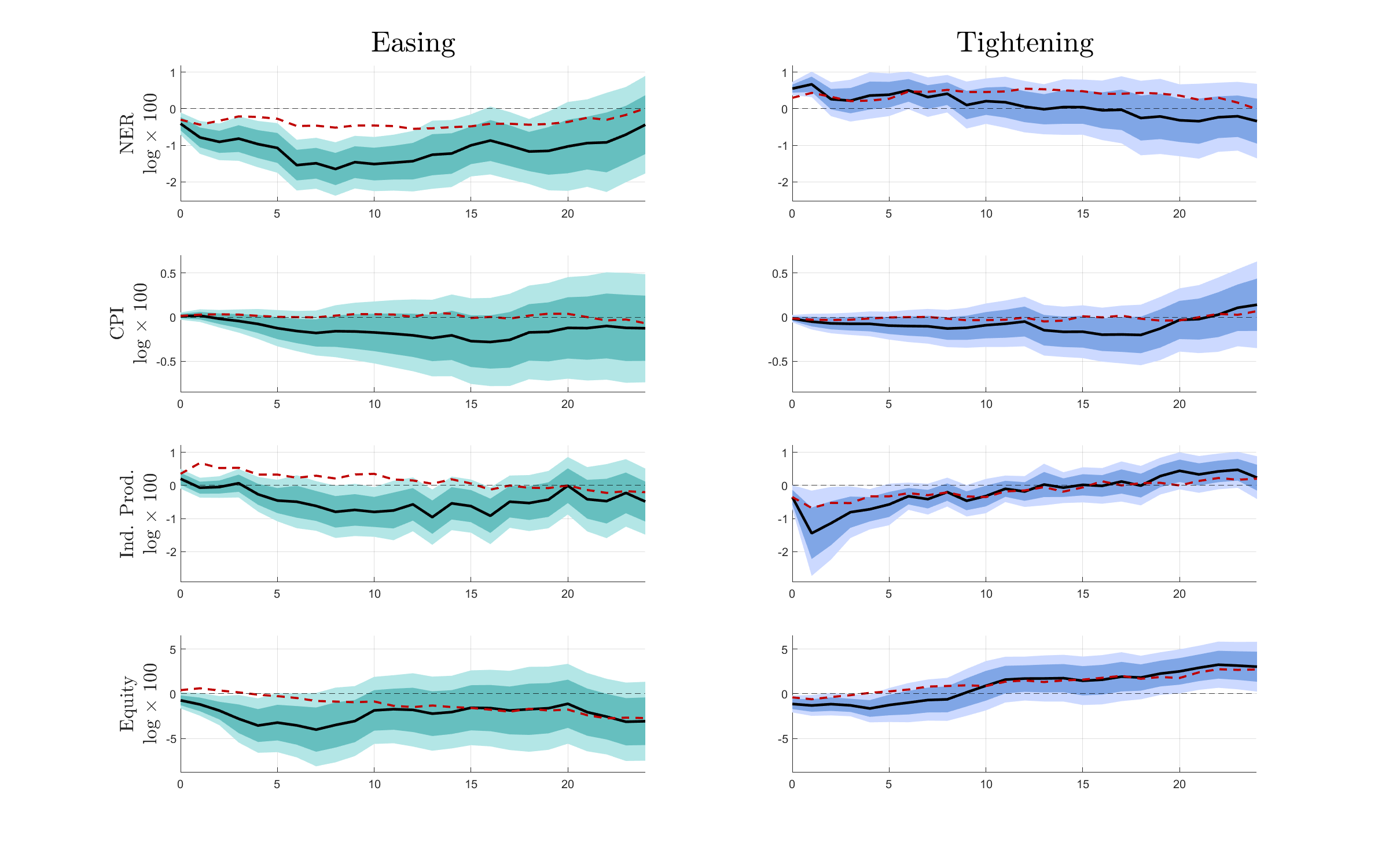}
\caption{Sign-Conditioned Spillovers of ECB Monetary Policy Shocks: Interacted Controls.}
\label{fig:ecb_signconditioned}
\floatfoot{\footnotesize
\textbf{Notes:} This figure reports impulse responses to ECB monetary policy shocks estimated using sign-conditioned local projections with interacted euro area controls. The specification allows both the shock and the propagation of lagged macroeconomic conditions to differ across tightening and easing/non-positive episodes. Solid black lines denote point estimates. Dark (light) shaded areas correspond to 68\% (90\%) confidence intervals. The red dotted line overlays the symmetric benchmark response (sign-flipped in the easing column). Responses are reported over a 24-month horizon, and standard errors are clustered by date.}
\end{figure}

Figures~\ref{fig:fed_signconditioned} and~\ref{fig:ecb_signconditioned} present the resulting impulse responses for Federal Reserve and ECB monetary policy shocks. Despite the additional flexibility introduced by interacting the entire set of controls with the sign of the shock, the qualitative patterns documented in the benchmark analysis remain unchanged. Tightening shocks generate large and persistent spillovers across exchange rates, prices, real activity, and equity markets, while easing (non-positive) shocks produce markedly weaker responses. If anything, the gap between tightenings and easings becomes more transparent under this specification, consistent with the view that asymmetry reflects genuinely different propagation dynamics rather than a restrictive baseline functional form.

%%%%%%%%%%%%%%%%%%%%%%%%%%%%%%%%%%%%%%%%%%%%%%%%%%%%%%%%%%%%%%%%%%%
\subsection{Robustness to Specification Choices}
\label{subsec:spec_robustness}

\begin{figure}[htbp]
\centering
\includegraphics[width=0.95\textwidth]{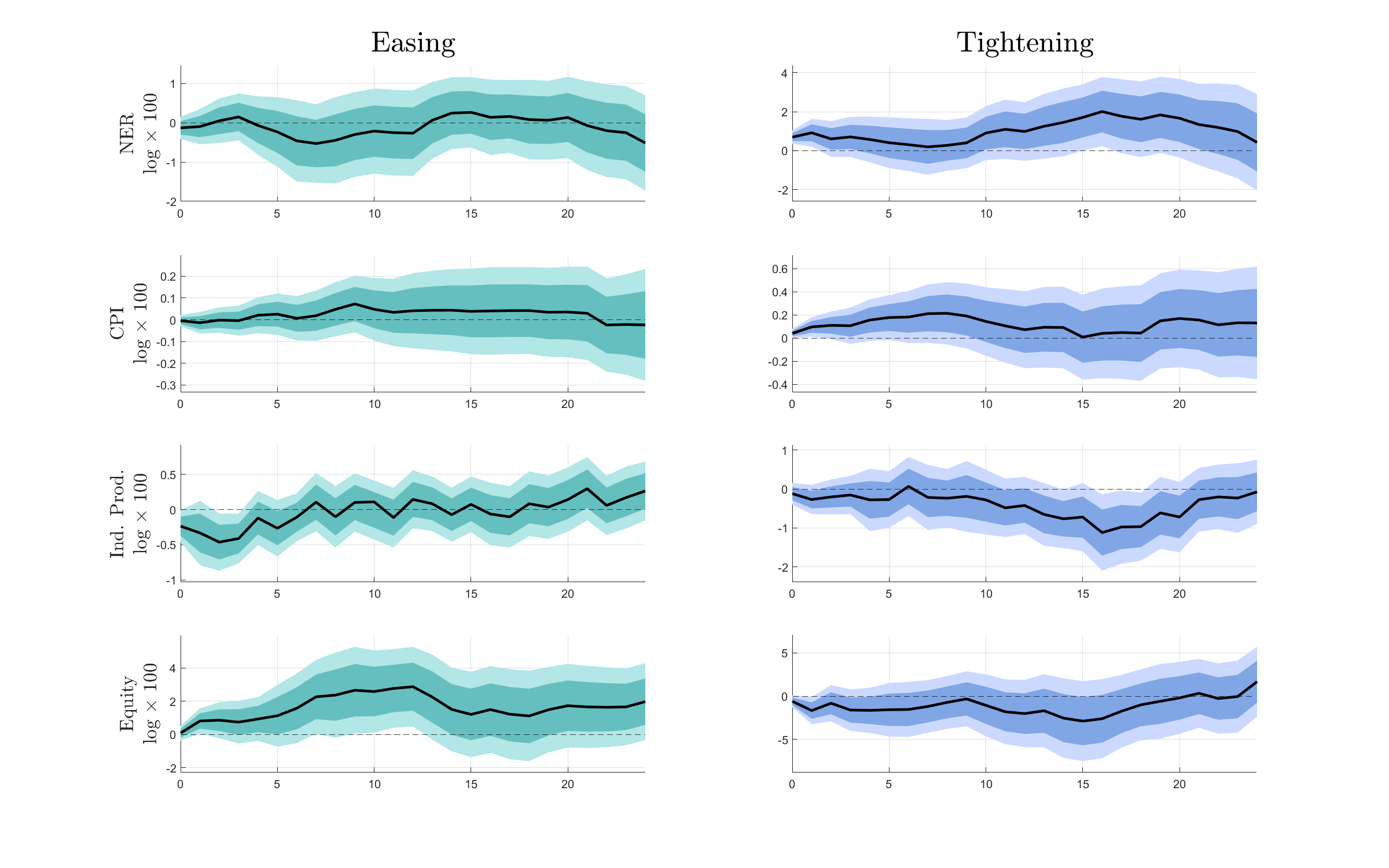}
\caption{Sign-dependent spillovers of U.S. monetary policy shocks with three lags.
The figure reports local projection impulse responses to Federal Reserve monetary policy shocks,
distinguishing between easing and tightening episodes. The red dotted line corresponds to the
symmetric benchmark specification. Shaded areas denote 68\% and 90\% confidence intervals.
Horizons are shown from 0 to 24 months.}
\label{fig:fed_l3}
\end{figure}

\begin{figure}[htbp]
\centering
\includegraphics[width=0.95\textwidth]{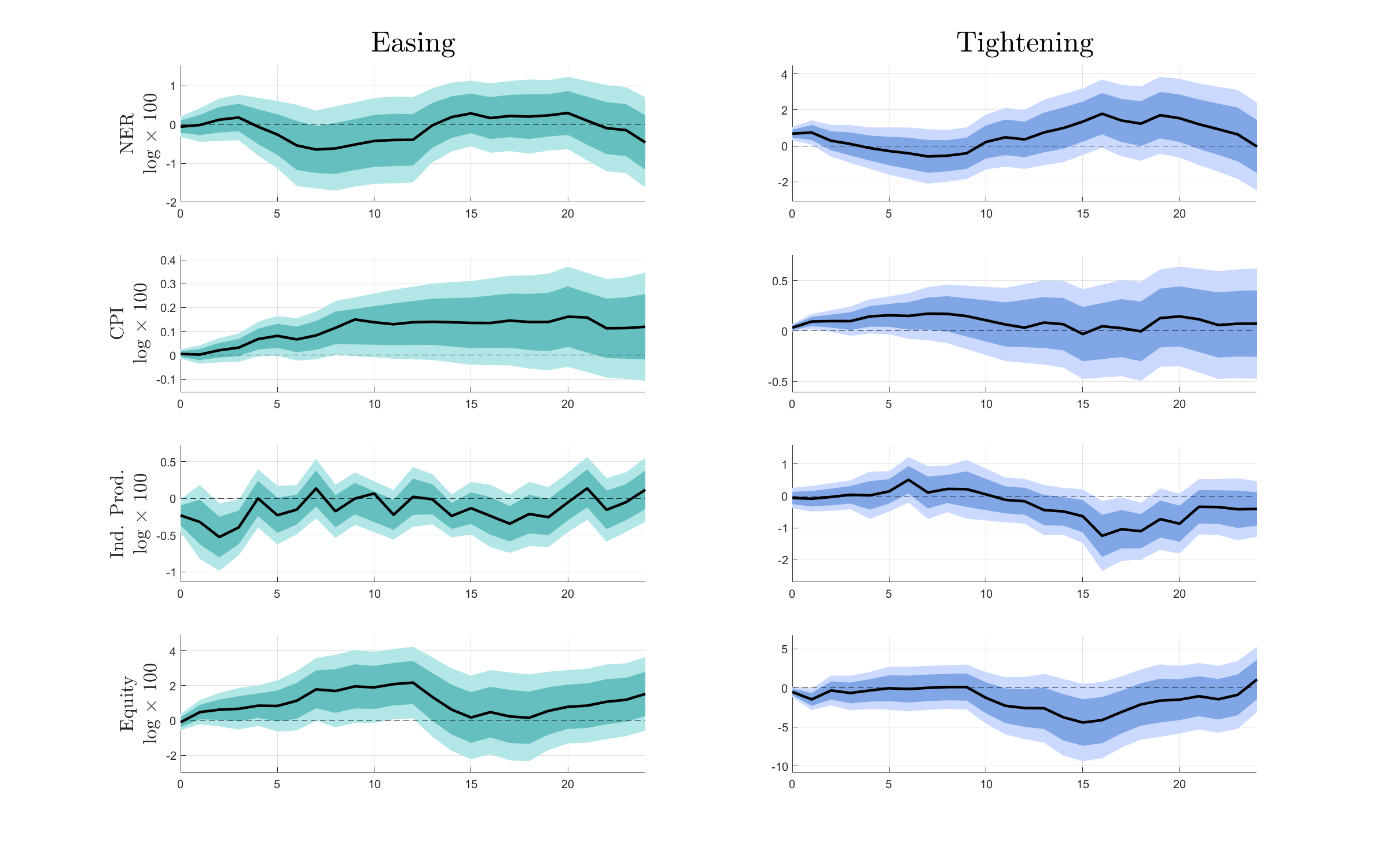}
\caption{Sign-dependent spillovers of U.S. monetary policy shocks with three lags.
The figure reports local projection impulse responses to Federal Reserve monetary policy shocks,
distinguishing between easing and tightening episodes. The red dotted line corresponds to the
symmetric benchmark specification. Shaded areas denote 68\% and 90\% confidence intervals.
Horizons are shown from 0 to 24 months.}
\label{fig:fed_l6}
\end{figure}

\begin{figure}[htbp]
\centering
\includegraphics[width=0.95\textwidth]{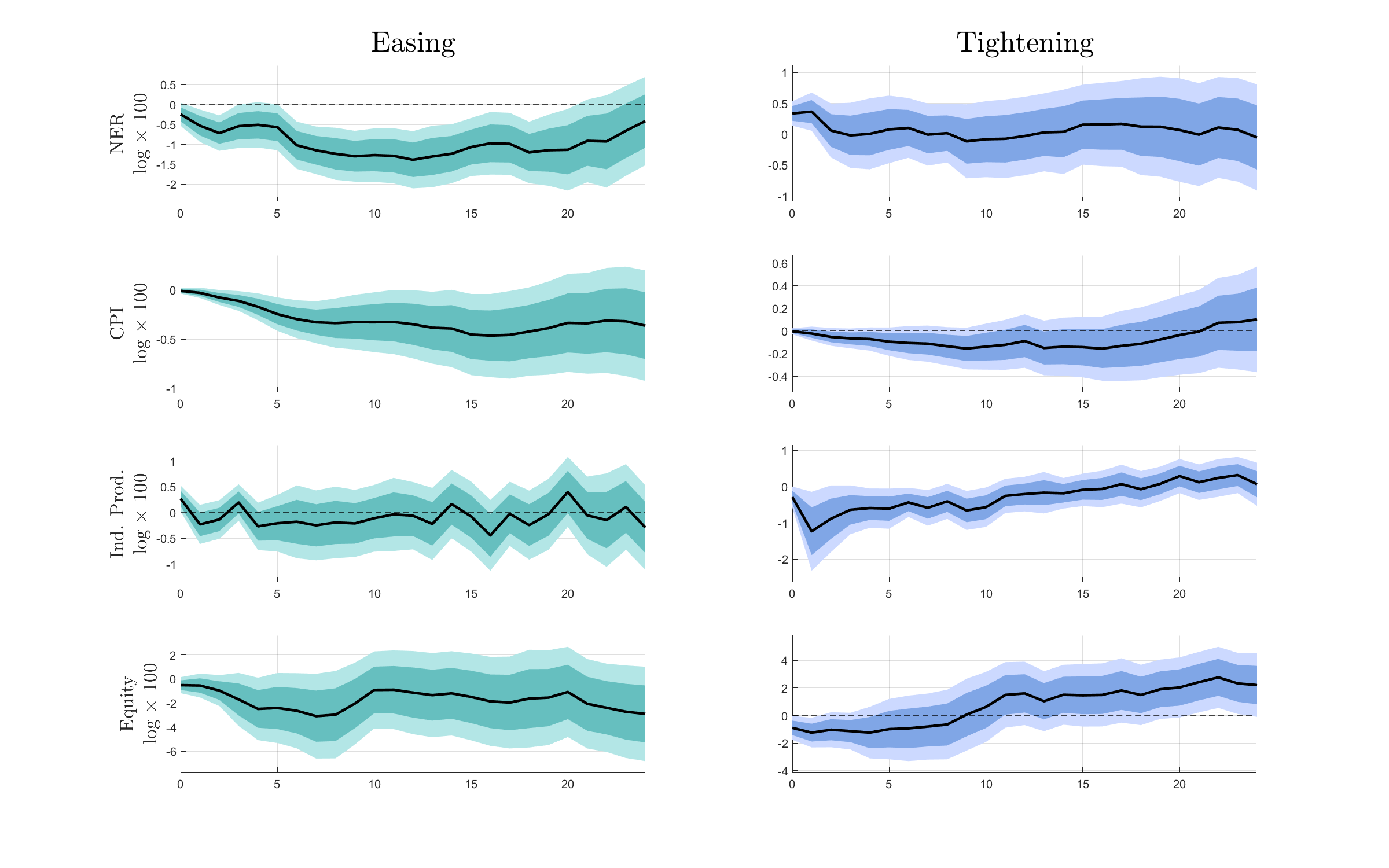}
\caption{Sign-dependent spillovers of ECB monetary policy shocks with three lags.
The figure shows impulse responses to ECB easing and tightening shocks, with the symmetric
benchmark overlaid. Results remain qualitatively unchanged relative to the baseline specification.}
\label{fig:ecb_l3}
\end{figure}

\begin{figure}[htbp]
\centering
\includegraphics[width=0.95\textwidth]{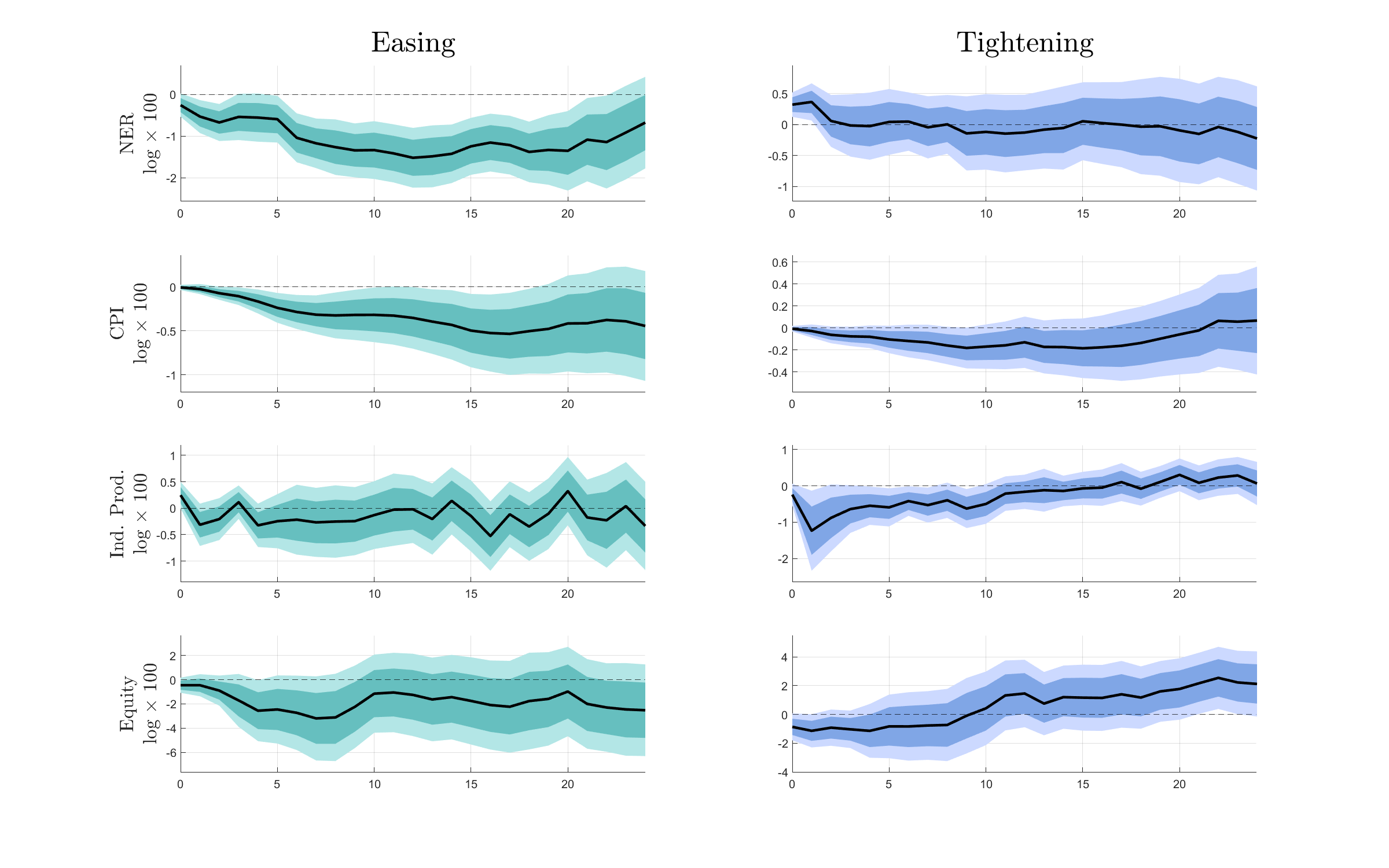}
\caption{Sign-dependent spillovers of ECB monetary policy shocks with six lags.
Increasing the number of lags does not alter the main qualitative patterns: tightening shocks
continue to drive the bulk of the international transmission, while easing shocks generate more
muted responses.}
\label{fig:ecb_l6}
\end{figure}

\paragraph{Alternative lag structures.}
We re-estimate the local projections using three and six lags. Figures~\ref{fig:fed_l3}--\ref{fig:ecb_l6} show that the magnitude, persistence, and asymmetry of international spillovers remain remarkably stable across lag specifications.

\begin{figure}[!htbp]
\centering
\caption{Federal Reserve Spillovers without Fixed Effects}
\label{fig:fed_nofe}
\includegraphics[width=0.95\textwidth]{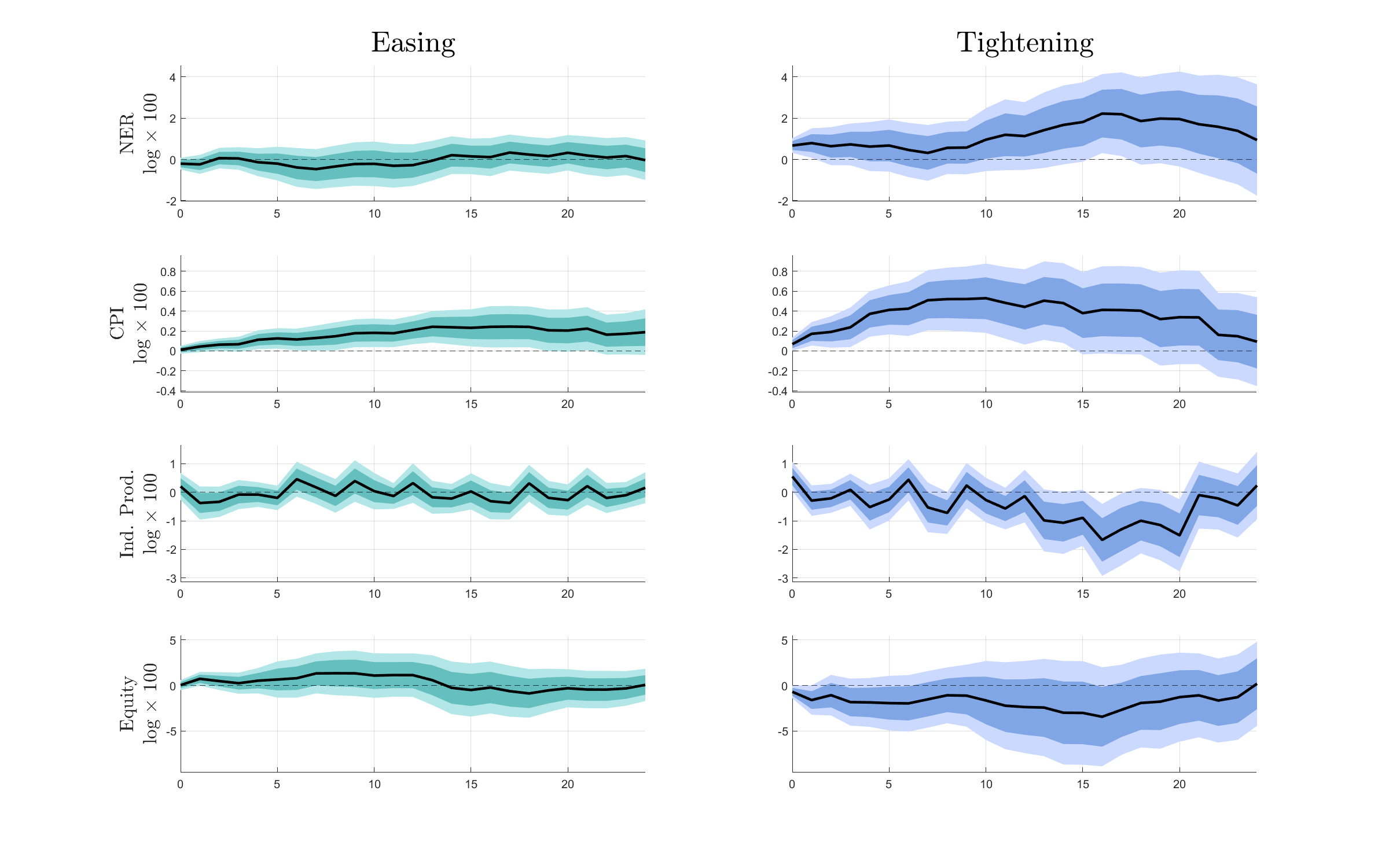}
\floatfoot{\footnotesize
Impulse responses to Federal Reserve monetary policy shocks estimated using local projections without country–month fixed effects.
The left column reports responses to easing shocks, while the right column reports responses to tightening shocks.
Shaded areas denote 68 and 90 percent confidence intervals.
The dashed line represents the symmetric benchmark response.
}
\end{figure}

\begin{figure}[!htbp]
\centering
\caption{ECB Spillovers without Fixed Effects}
\label{fig:ecb_nofe}
\includegraphics[width=0.95\textwidth]{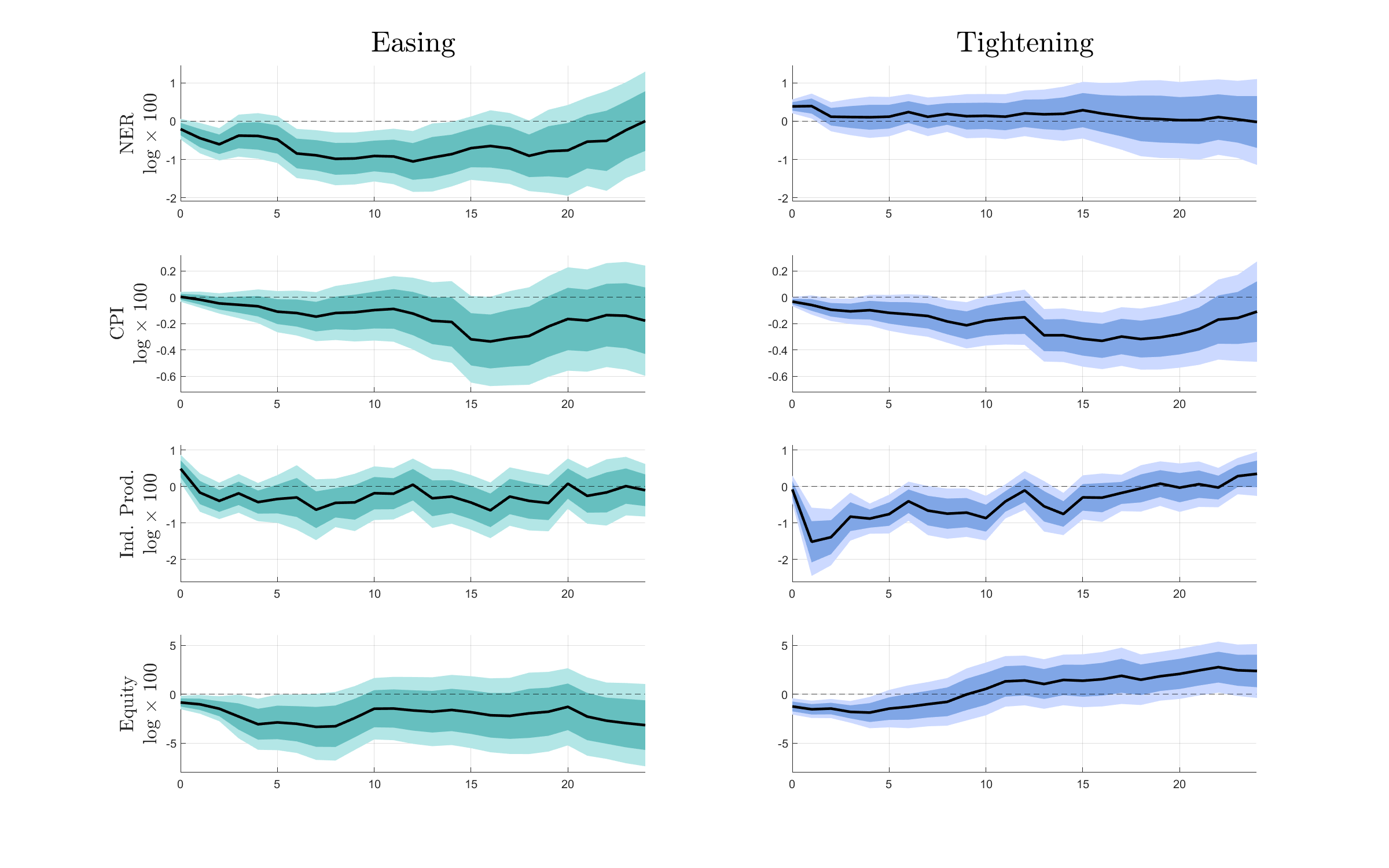}
\floatfoot{\footnotesize
Impulse responses to ECB monetary policy shocks estimated using local projections without country–month fixed effects.
The left column reports responses to easing shocks, while the right column reports responses to tightening shocks.
Shaded areas denote 68 and 90 percent confidence intervals.
The dashed line represents the symmetric benchmark response.
}
\end{figure}

\paragraph{Removing fixed effects.}
We next re-estimate the local projections without country–month fixed effects. Figures~\ref{fig:fed_nofe} and~\ref{fig:ecb_nofe} show that the qualitative asymmetries remain intact, confirming that the results are not driven by high-dimensional fixed effects.

%%%%%%%%%%%%%%%%%%%%%%%%%%%%%%%%%%%%%%%%%%%%%%%%%%%%%%%%%%%%%%%
\subsection{Robustness: Sub-sample Analysis}
\label{subsec:sample_robustness}

\begin{figure}
\centering
\caption{Federal Reserve Shocks: Advanced vs Non-Advanced Economies}
\label{fig:fed_ae_noneae}

\begin{subfigure}{0.49\textwidth}
    \centering
    \includegraphics[width=\textwidth]{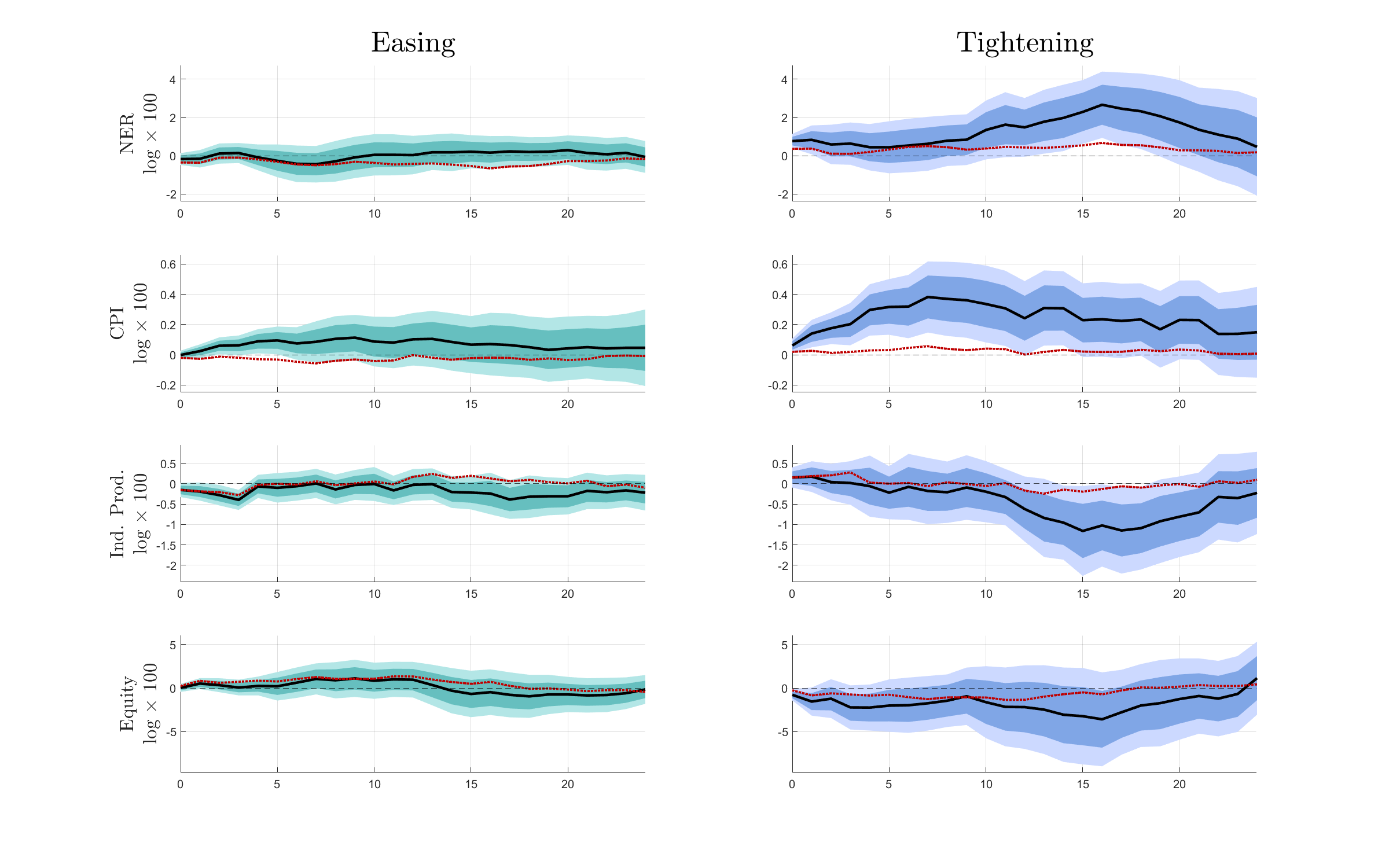}
    \caption{Advanced Economies}
\end{subfigure}
\hfill
\begin{subfigure}{0.49\textwidth}
    \centering
    \includegraphics[width=\textwidth]{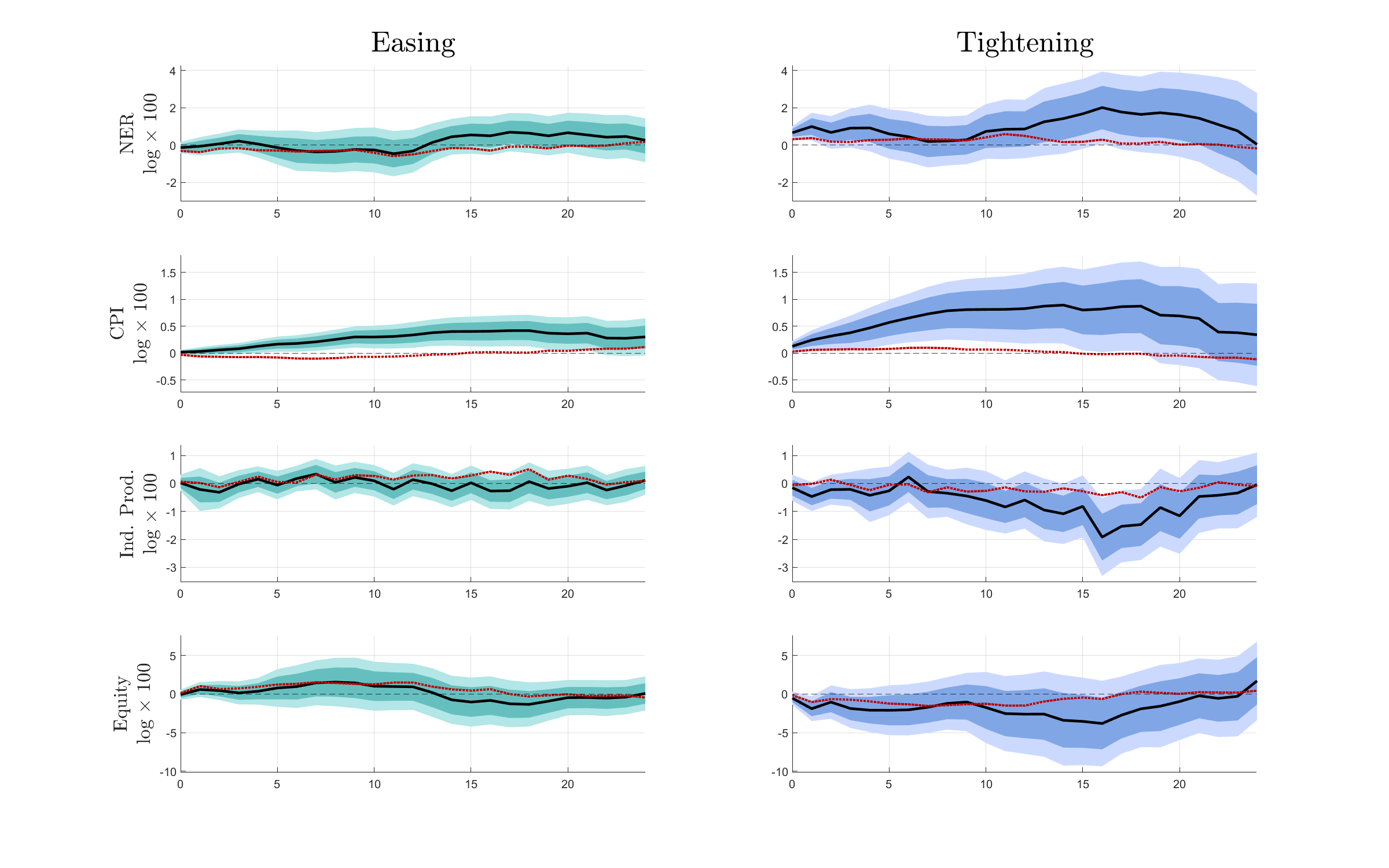}
    \caption{Non-Advanced Economies}
\end{subfigure}

\medskip
\footnotesize
\textit{Notes:} The figure reports sign-dependent impulse responses to Federal Reserve monetary policy shocks estimated separately for Advanced and Non-Advanced Economies. Solid lines denote point estimates; shaded areas correspond to 68\% and 90\% confidence intervals. Responses are shown for a 24-month horizon.
\end{figure}

\begin{figure}
\centering
\caption{ECB Shocks: Advanced vs Non-Advanced Economies}
\label{fig:ecb_ae_noneae}

\begin{subfigure}{0.49\textwidth}
    \centering
    \includegraphics[width=\textwidth]{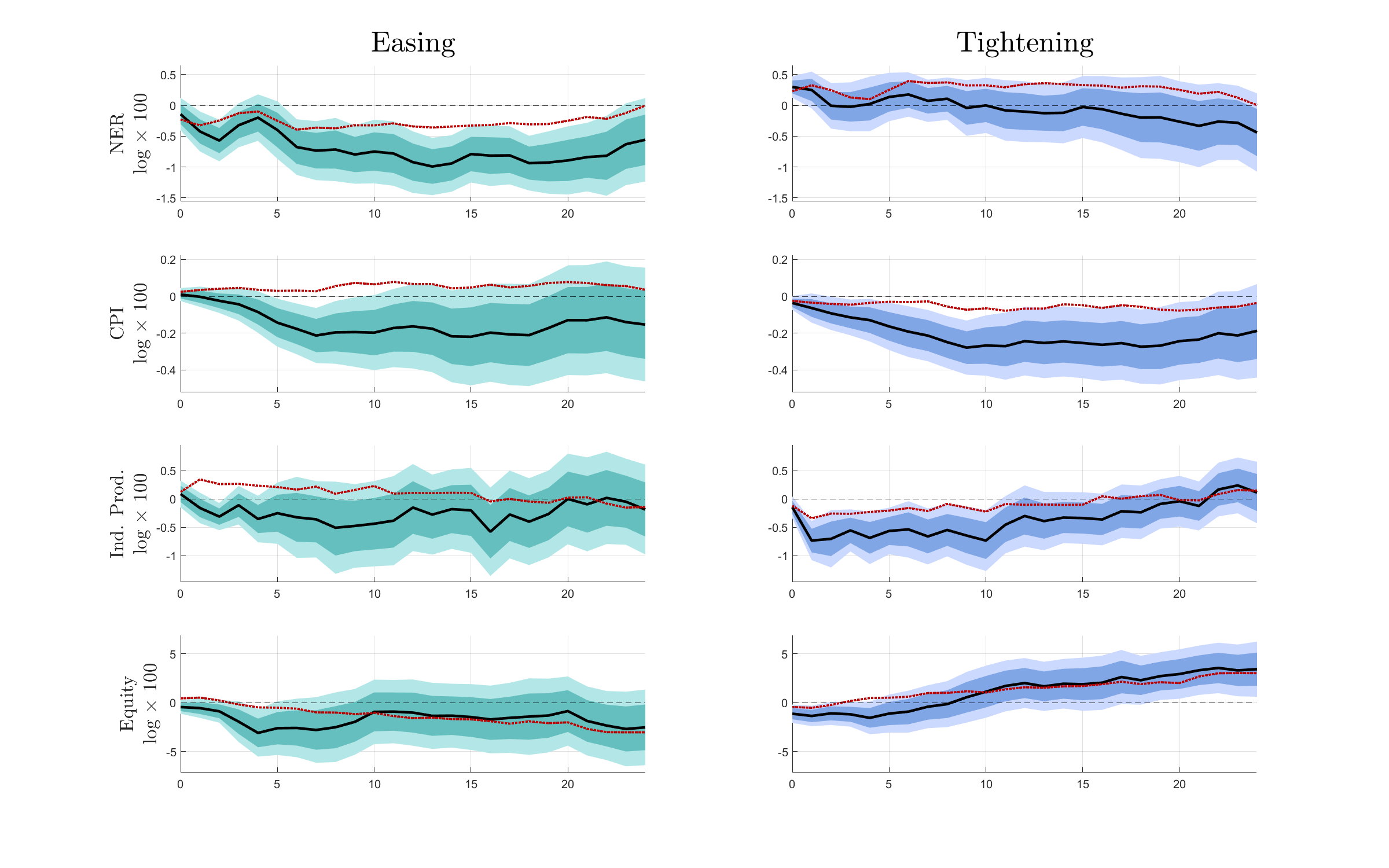}
    \caption{Advanced Economies}
\end{subfigure}
\hfill
\begin{subfigure}{0.49\textwidth}
    \centering
    \includegraphics[width=\textwidth]{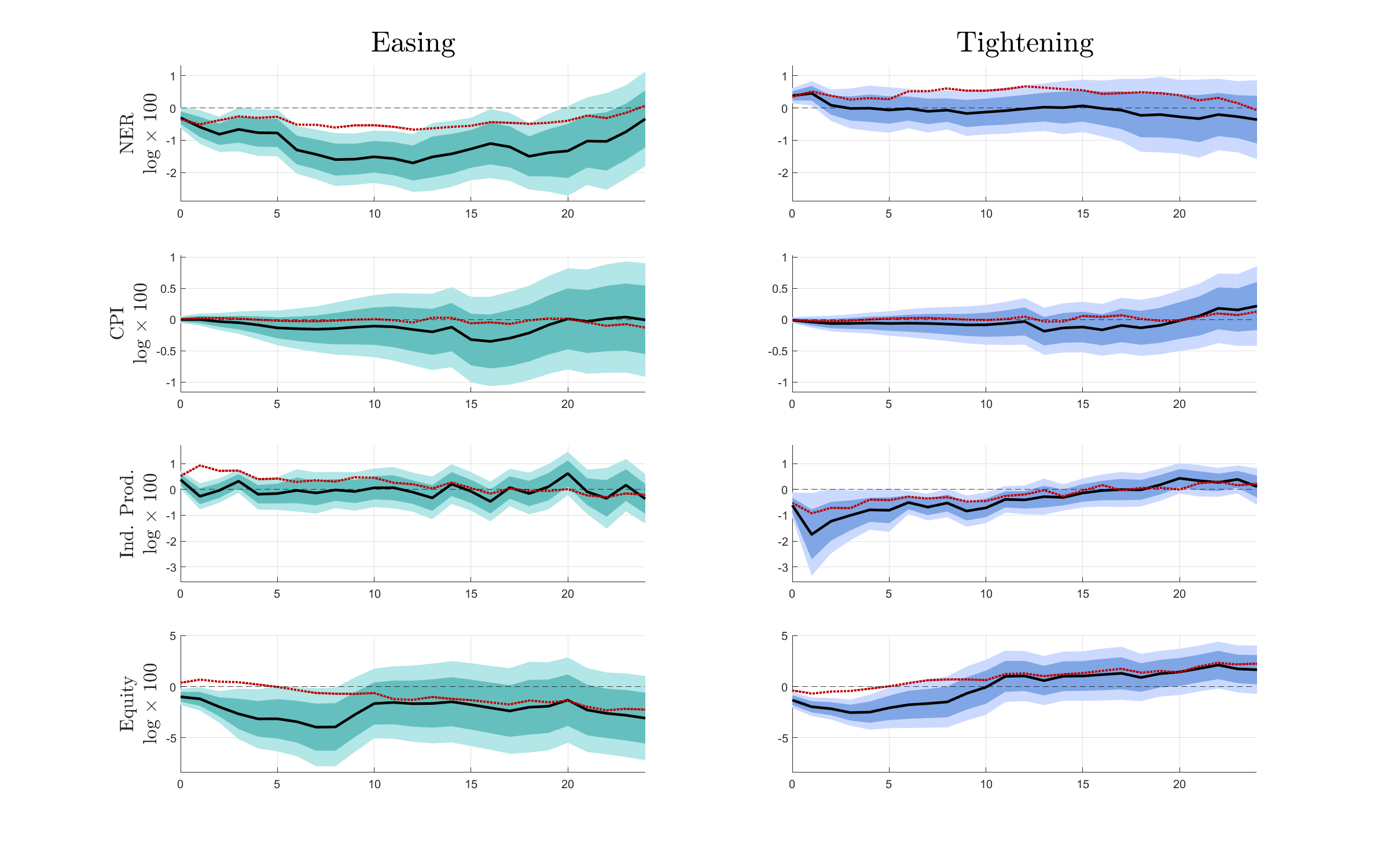}
    \caption{Non-Advanced Economies}
\end{subfigure}

\medskip
\footnotesize
\textit{Notes:} The figure reports sign-dependent impulse responses to ECB monetary policy shocks estimated separately for Advanced and Non-Advanced Economies. Solid lines denote point estimates; shaded areas correspond to 68\% and 90\% confidence intervals. Responses are shown for a 24-month horizon.
\end{figure}

\paragraph{Advanced vs Non-Advanced Economies.}
Figures \ref{fig:fed_ae_noneae} and \ref{fig:ecb_ae_noneae} report results from splitting the sample into Advanced and Non-Advanced Economies. The estimated impulse responses display highly similar sign-dependent patterns across the two groups. In both cases, international spillovers are overwhelmingly driven by tightening shocks, with larger magnitudes in Non-Advanced Economies.

\begin{figure}[htbp]
    \centering
    \includegraphics[width=0.95\textwidth]{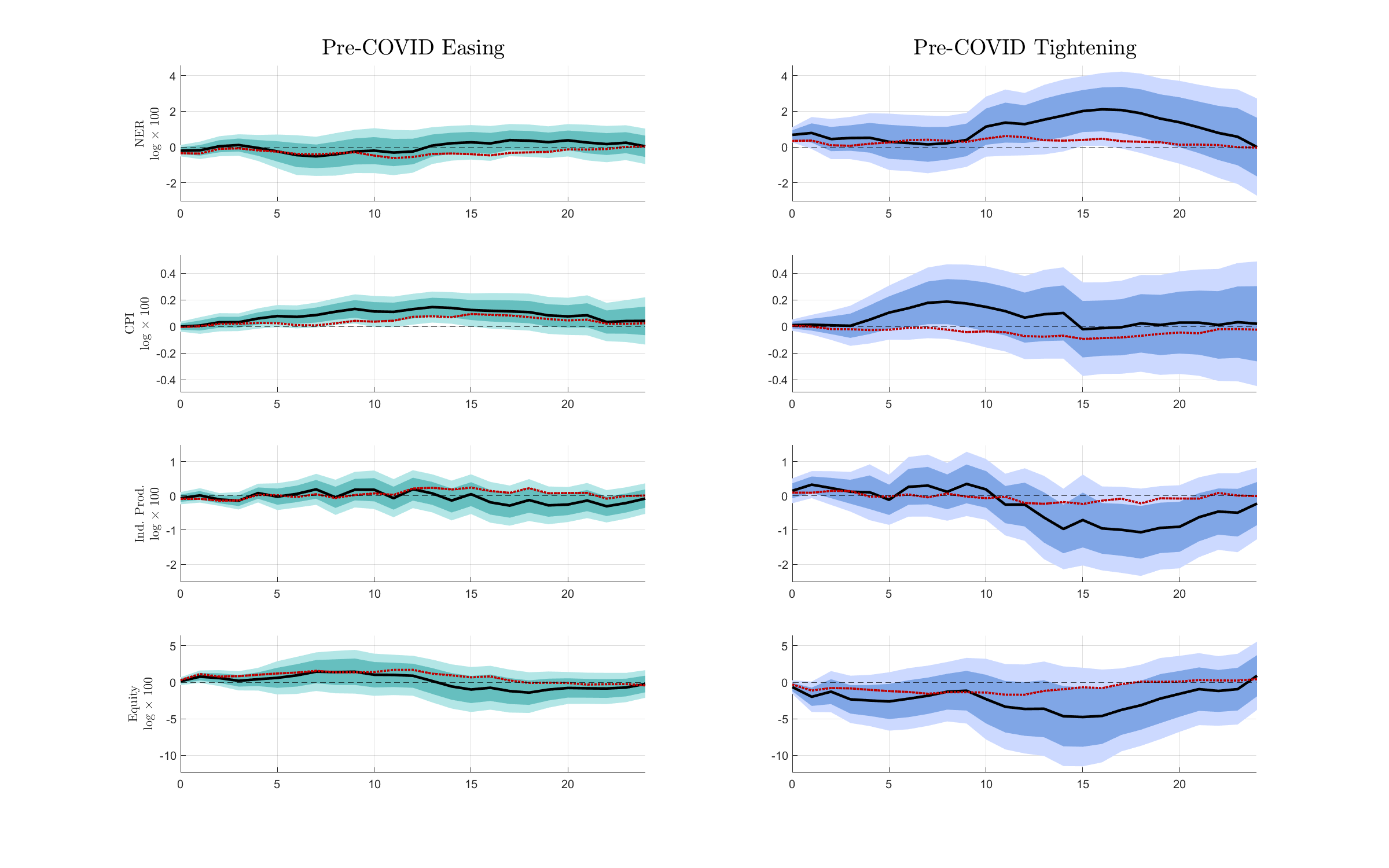}
    \caption{Pre-COVID sample: Sign-dependent impulse responses to Federal Reserve monetary policy shocks. The figure reports impulse responses to easing (left column) and tightening (right column) shocks. Shaded areas denote 68\% and 90\% confidence intervals.}
    \label{fig:fed_precovid}
\end{figure}

\begin{figure}[htbp]
    \centering
    \includegraphics[width=0.95\textwidth]{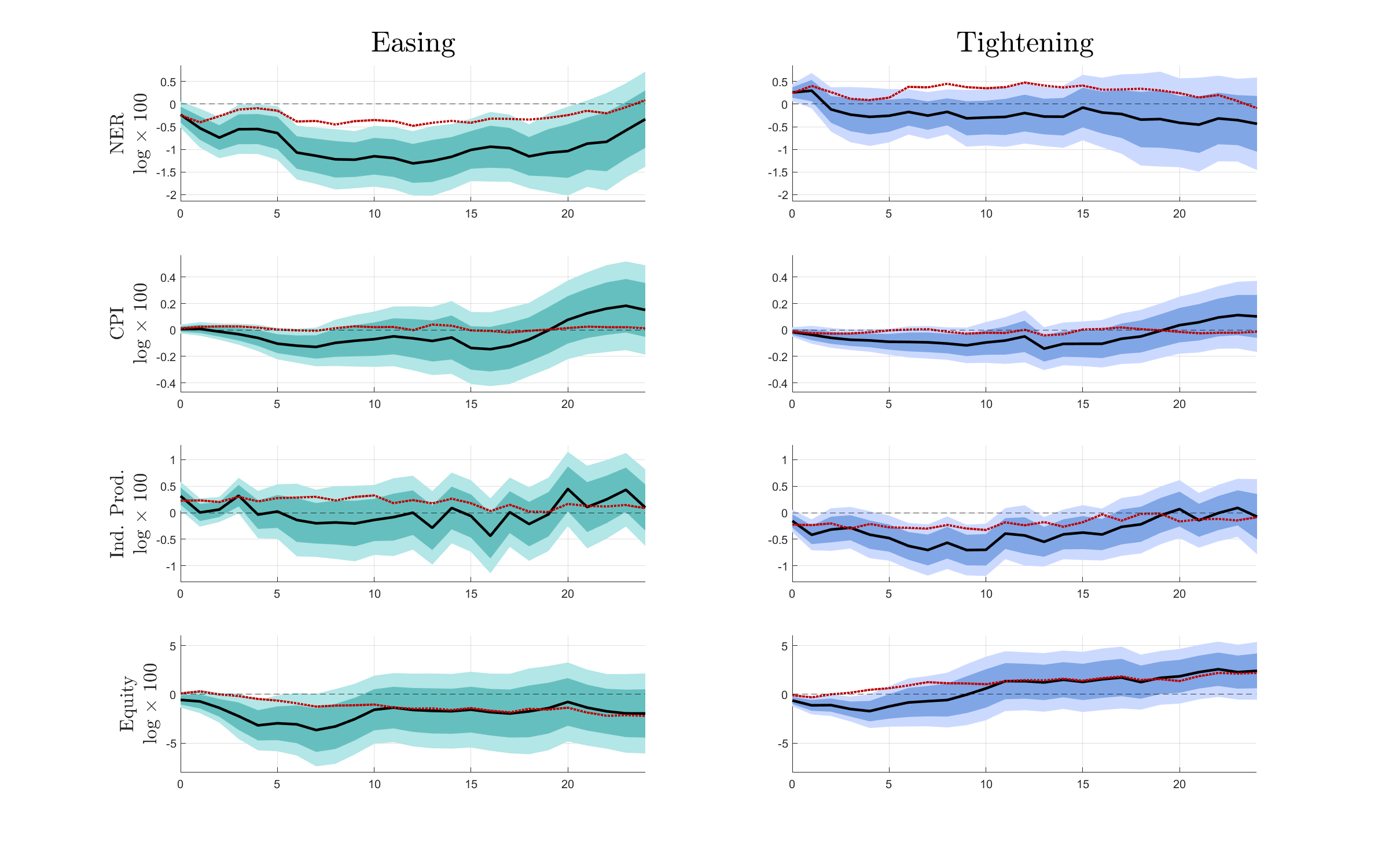}
    \caption{Pre-COVID sample: Sign-dependent impulse responses to ECB monetary policy shocks. The figure reports impulse responses to easing (left column) and tightening (right column) shocks. Shaded areas denote 68\% and 90\% confidence intervals.}
    \label{fig:ecb_precovid}
\end{figure}

\paragraph{Pre-COVID sample.}
Figures \ref{fig:fed_precovid} and \ref{fig:ecb_precovid} report results obtained by restricting the sample to the pre-COVID period. The main findings remain unchanged: contractionary monetary policy shocks generate larger and more persistent international spillovers than expansionary shocks for both central banks.

\begin{figure}[htbp]
    \centering
    \caption{Federal Reserve Monetary Policy Shocks: Post-GFC Sample}
    \label{fig:fed_postgfc}
    \includegraphics[width=0.95\textwidth]{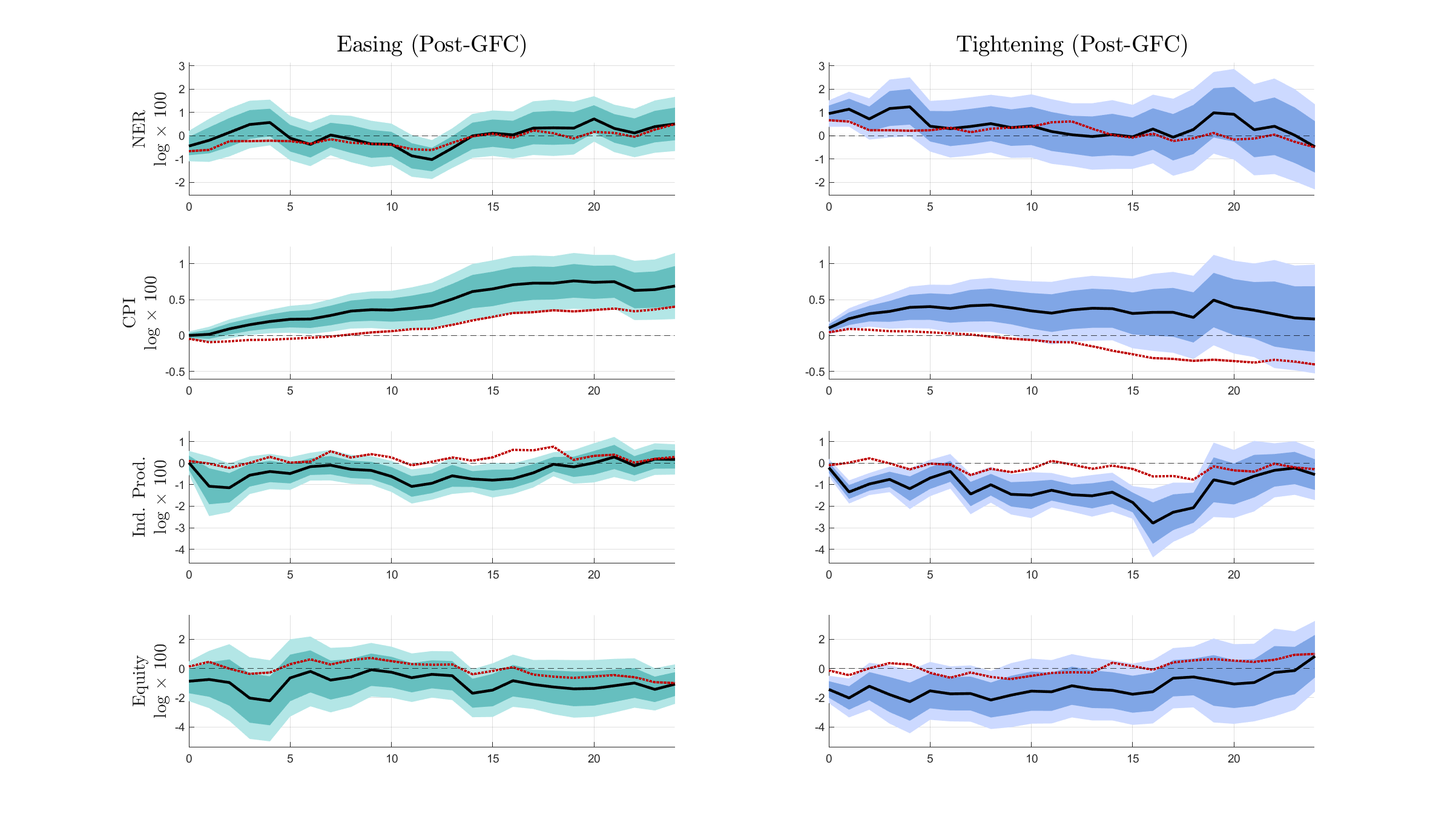}
    \floatfoot{\footnotesize
    Notes: The figure reports impulse responses to Federal Reserve monetary policy shocks estimated on the post–Global Financial Crisis sample. The left column shows easing shocks and the right column tightening shocks. Solid lines denote point estimates, while shaded areas correspond to 68\% and 90\% confidence intervals.}
\end{figure}

\begin{figure}[htbp]
    \centering
    \caption{European Central Bank Monetary Policy Shocks: Post-GFC Sample}
    \label{fig:ecb_postgfc}
    \includegraphics[width=0.95\textwidth]{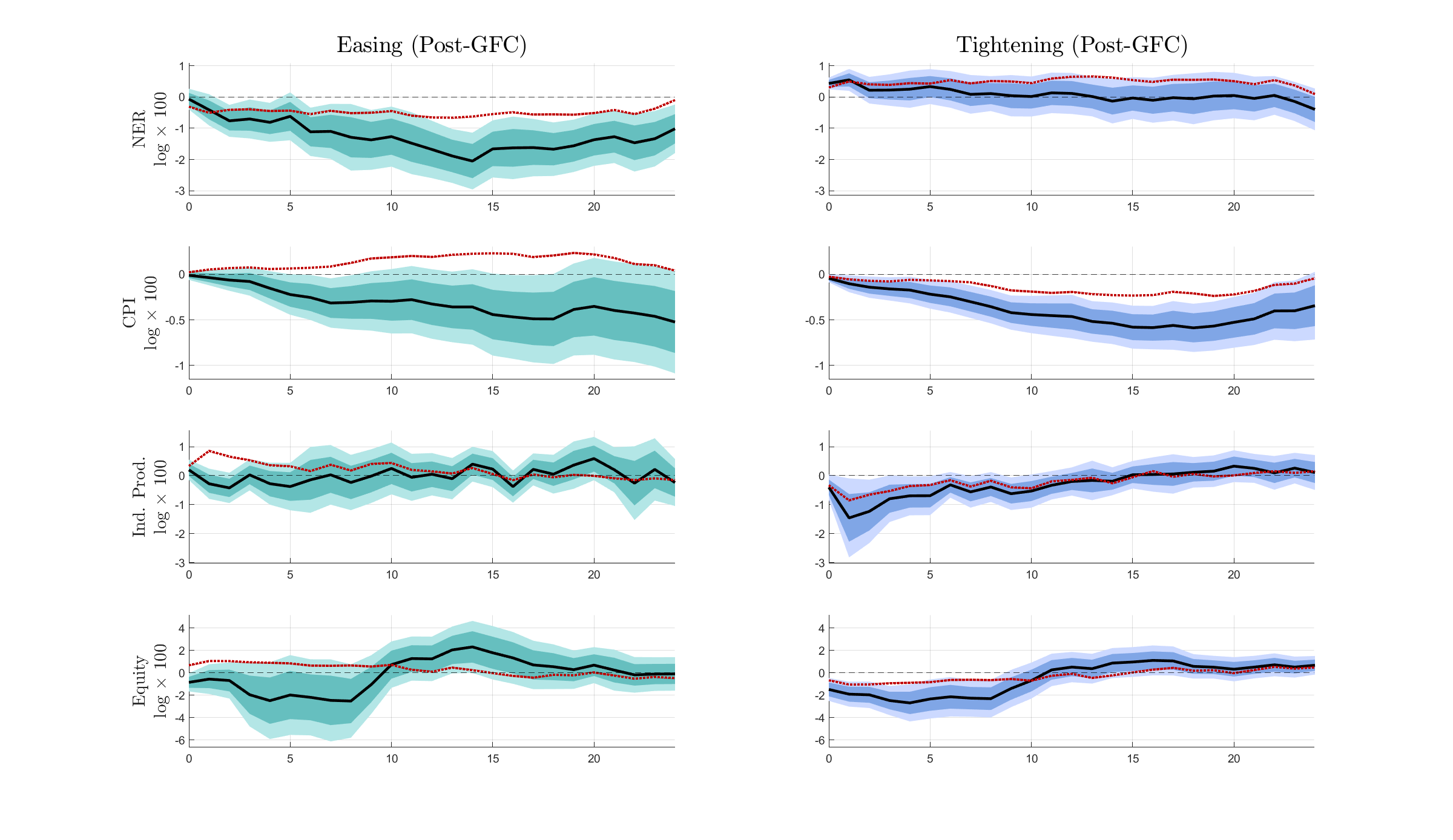}
    \floatfoot{\footnotesize
    Notes: The figure reports impulse responses to European Central Bank monetary policy shocks estimated on the post–Global Financial Crisis sample. The left column shows easing shocks and the right column tightening shocks. Solid lines denote point estimates, while shaded areas correspond to 68\% and 90\% confidence intervals.}
\end{figure}

\paragraph{Post-GFC sample.}
Figures \ref{fig:fed_postgfc} and \ref{fig:ecb_postgfc} report results for the post–Global Financial Crisis sample. The sign-dependent patterns are highly similar to those in the full sample, confirming that asymmetric international monetary transmission is a defining feature of the modern global financial environment.

%%%%%%%%%%%%%%%%%%%%%%%%%%%%%%%%%%%%%%%%%%%%%%%%%%%%%%%%%%%%%%%%%%%%%%
\section{Conclusions} \label{sec:conclusions}

This paper studies the international spillovers of Federal Reserve and European Central Bank monetary policy shocks through the lens of sign dependence. Using a unified high-frequency identification of pure monetary policy shocks and a parsimonious sign-dependent local projection framework, we show that international monetary spillovers are overwhelmingly driven by contractionary policy episodes. For both central banks, monetary tightenings generate large and persistent deteriorations in global financial conditions, contractions in real activity, and sharp declines in international trade. By contrast, expansionary monetary policy shocks do not produce a symmetric reversal of these effects.

A central finding of the paper is that linear empirical models provide a misleading characterization of international monetary transmission. By imposing symmetry between tightening and easing shocks, linear specifications average strong contractionary spillovers with weak or countervailing responses to expansionary shocks, substantially attenuating estimated effects and, in some cases, obscuring their sign. Allowing for sign dependence reveals that the international propagation of monetary policy is not well approximated by a symmetric mechanism and that contractionary shocks account for virtually all economically meaningful spillovers.

Beyond documenting asymmetric spillovers, the paper highlights an important empirical puzzle: expansionary euro area monetary policy shocks are followed by a deterioration in global financial conditions and international trade. Our evidence suggests that this pattern reflects endogenous policy interactions and global financial dominance, whereby ECB easings trigger a tightening of U.S. monetary policy and dollar-denominated financial conditions. While the paper does not seek to provide a structural explanation for this mechanism, it documents a robust empirical regularity that future theoretical work will need to confront.

Overall, the results underscore the importance of accounting for sign dependence when measuring international monetary spillovers. From an econometric perspective, ignoring asymmetries leads to systematic understatement of the global effects of monetary policy tightenings. From a policy perspective, the findings suggest that contractionary policy actions by major central banks carry disproportionately large international consequences, while expansionary actions do not deliver commensurate relief abroad. Accurately measuring these asymmetries is therefore essential for understanding the global transmission of monetary policy and for informing policy discussions in an increasingly interconnected financial system.

\newpage
\printbibliography

%\newpage
%\section{Main Manuscript Figures} \label{sec:main_manuscript_figures}

%%%%%%%%%%%%%%%%%%%%%%%%%%%%%%%%%%%%%%%%%%%%%%%%%%%%%%%%%%%%%%%%%%%%%%
% Appendix
%%%%%%%%%%%%%%%%%%%%%%%%%%%%%%%%%%%%%%%%%%%%%%%%%%%%%%%%%%%%%%%%%%%%%%%%%%%%%%%%%%%%%%

\clearpage
\appendix

%%%%%%%%%%%%%%%%%%%%%%%%%%%%%%%%%%%%%%%%%%%%%%%%%%%%%%%%%%%%%%%%%%%%%%%%%%%%%%%%%%%%%%
\newpage
\section{Data Details} \label{sec:appendix_data_details}

\begin{table}[ht]
\centering
\caption{Advanced Economies in the Sample}
\small
\label{tab:sample_advanced}
\begin{tabular}{lc}
\toprule
Country & Number of Months \\
\midrule
Australia        & 224 \\
Canada           & 220 \\
Croatia          & 288 \\
Czech Republic   & 305 \\
Denmark          & 305 \\
Estonia          & 144 \\
Iceland          & 230 \\
Israel           & 300 \\
Japan            & 220 \\
Korea            & 304 \\
Latvia           & 168 \\
Lithuania        & 181 \\
Norway           & 305 \\
Singapore        & 270 \\
Slovak Republic  & 120 \\
Slovenia         & 98  \\
Sweden           & 220 \\
Switzerland      & 162 \\
United Kingdom   & 303 \\
\bottomrule
\end{tabular}
\footnotesize
\floatfoot{\textit{Notes:} The table reports the non-advanced economies included in the sample and the number of monthly observations available for each country. The panel is unbalanced due to differences in data availability across countries.}
\end{table}

\begin{table}[ht]
\centering
\caption{Non-Advanced Economies in the Sample}
\label{tab:sample_non_advanced}
\small
\begin{tabular}{lc}
\toprule
Country & Number of Months \\
\midrule
Bangladesh        & 271 \\
Bosnia and Herzegovina & 192 \\
Brazil            & 305 \\
Bulgaria          & 283 \\
Chile             & 303 \\
China (Mainland)  & 88  \\
Colombia          & 302 \\
Hungary           & 305 \\
India             & 220 \\
Malaysia          & 243 \\
Mexico            & 274 \\
Mongolia          & 189 \\
Montenegro        & 226 \\
North Macedonia   & 228 \\
Pakistan          & 220 \\
Peru              & 232 \\
Philippines       & 288 \\
Poland            & 305 \\
Romania           & 301 \\
Russia            & 220 \\
Türkiye           & 305 \\
Ukraine           & 264 \\
Uzbekistan        & 82  \\
Venezuela         & 109 \\
Vietnam           & 112 \\
\bottomrule
\end{tabular}
\footnotesize
\floatfoot{\textit{Notes:} The table reports the non-advanced economies included in the sample and the number of monthly observations available for each country. The panel is unbalanced due to differences in data availability across countries.}
\end{table}

%%%%%%%%%%%%%%%%%%%%%%%%%%%%%%%%%%%%%%%%%%%%%%%%%%%%%%%%%%%%%%%%%%%%%%%%%%%%%%%%%%%%%%
\clearpage
\newpage
\section{Additional Results Figures} \label{sec:appendix_additional_results}

\end{document}